\documentclass[aps,prd,twocolumn,nofootinbib,superscriptaddress,10pt]{revtex4-1}

\usepackage{amsmath,amsthm,latexsym,amssymb,amsfonts}
\usepackage{graphicx,color}
\usepackage{hyperref,natbib}
\usepackage[utf8]{inputenc}
\usepackage[T1]{fontenc}
\usepackage{textcomp}
\usepackage[cal=boondox]{mathalfa}
\def\bf{\textbf}

\begin{document}

\title{Hamiltonian formalism for $f(T)$ gravity}

\author{Rafael Ferraro}
\email[Member of Carrera del Investigador Cient\'{\i}fico (CONICET,
Argentina); ]{ferraro@iafe.uba.ar}\affiliation{Instituto de
Astronom\'\i a y F\'\i sica del Espacio (IAFE, CONICET-UBA), Casilla
de Correo 67, Sucursal 28, 1428 Buenos Aires, Argentina.}
\affiliation{Departamento de F\'\i sica, Facultad de Ciencias
Exactas y Naturales, Universidad de Buenos Aires, Ciudad
Universitaria, Pabell\'on I, 1428 Buenos Aires, Argentina.}

\author{Mar\'ia Jos\'e Guzm\'an}
\email[]{mjguzman@iafe.uba.ar}\affiliation{Instituto de Astronom\'\i
a y F\'\i sica del Espacio (IAFE, CONICET-UBA), Casilla de Correo
67, Sucursal 28, 1428 Buenos Aires, Argentina.}

\begin{abstract}
{\ We present the Hamiltonian formalism for $f(T)$ gravity, and prove that the theory has $\frac{n(n-3)}{2}+1$ degrees of freedom (d.o.f.) in $n$ dimensions. We start from a scalar-tensor action for the theory, which represents a scalar field minimally coupled with the torsion scalar $T$ that defines the teleparallel equivalent of general relativity (TEGR) Lagrangian. $T$ is written as a quadratic form of the coefficients of anholonomy of the vierbein. We obtain the primary constraints through the analysis of the structure of the eigenvalues of the multi-index matrix involved in the definition of the canonical momenta. The auxiliary scalar field generates one extra primary constraint when compared with the TEGR case. The secondary constraints are the super-Hamiltonian and supermomenta constraints, that are preserved from the Arnowitt-Deser-Misner formulation of GR. There is a set of $\frac{n(n-1)}{2}$ primary constraints that represent the local Lorentz transformations of the theory, which can be combined to form a set of $\frac{n(n-1)}{2}-1$ first-class constraints, while one of them becomes second-class. This result is irrespective of the dimension, due to the structure of the matrix of the brackets between the constraints. The first-class canonical Hamiltonian is modified due to this local Lorentz violation, and the only one local Lorentz transformation that becomes second-class pairs up with the second-class constraint $\pi \approx 0$ to remove one d.o.f. from the $n^2+1$ pairs of canonical variables. The remaining $\frac{n(n-1)}{2}+2n-1$ primary constraints remove the same number of d.o.f., leaving the theory with $\frac{n(n-3)}{2}+1$ d.o.f. This means that $f(T)$ gravity has only one extra d.o.f., which could be interpreted as a scalar d.o.f.
}
\end{abstract}

\maketitle

\section{Introduction}

The inspiration for the development of modified theories of gravity is intimately related with problems appearing in the realm of cosmology: the hypothesis of dark matter, the accelerated expansion of the universe, the inflation paradigm, among others. The emergence of singularities, together with the quest for a quantum field theory of gravity, is also a strong motivation for the study of deformations of general relativity (GR) in the high-energy regime.
There are many paths that lead to modified gravity, but we will focus on those that modify the Lagrangian of general relativity by an arbitrary function of it: that is the well-known $f(R)$ paradigm, which consists in just including an arbitrary function of the Ricci scalar. This theory was the inspiration for $f(T)$ gravity, a class of theories that have been proposed more than a decade ago in the context of teleparallelism \textit{\`a la} Born-Infeld \cite{Ferraro:2006jd}. The general $f(T)$ gravity is a modification of the teleparallel equivalent of general relativity (TEGR), whose Lagrangian is linear in the torsion scalar $T$. The main dynamical variable of this theory is the tetrad or vierbein field (\textit{vielbein} in $n$ dimensions), a field of orthonormal basis in the tangent space. The Lagrangian is quadratic in the torsion of the Weitzenb\"{o}ck connection, which is a curvatureless connection that defines a spacetime with absolute parallelism \cite{Weyl1918,Einstein1925,Einstein1928,Einstein1930a,Einstein1930b,Cartan1922a,Cartan1922b,Weitzenbock1923,Cho:1975dh}.  $f(T)$ theories have attracted a lot of attention \cite{Linder:2010py,Bamba:2010iw,Wu:2010xk,Bamba:2010wb,Yang:2010ji,Li:2010cg,Chen:2010va,Sotiriou:2010mv,Li:2011wu,Li:2011rn,Yang:2010ji,Izumi:2012qj,Fiorini:2013hva,Fiorini:2013kba,Chen:2014qtl,Fiorini:2015hob,Gonzalez:2015sha,Cai:2015emx,Fiorini:2016zrt,Bahamonde:2016kba,Bahamonde:2016grb,Golovnev:2017dox,Bahamonde:2017wwk,Mai:2017riq,Hohmann:2018rwf} since they can describe an inflationary expansion without resorting to an inflaton field \cite{Ferraro:2006jd,Ferraro:2008ey}. They can also produce an accelerated expansion at late times, so mimicking the effects of dark energy  \cite{Bengochea:2008gz}. Since the action of $f(T)$ gravity contains only first derivatives of the vierbein, the dynamical equations are always second order, which is also an appealing feature for any modified theory of gravity. In this respect $f(T)$ gravity separates from metric $f(R)$ gravity whose dynamical equations are fourth order.

$f(T)$ theories have also given rise to disputes about an intriguing and essential feature: the action of the theory is not invariant under local Lorentz transformations of the tetrad \cite{Ferraro:2006jd,Li:2010cg,Li:2011wu,Sotiriou:2010mv}. In fact, if a tetrad ${\bf E}_a$ solves the equations of motion, then it is not guaranteed that a locally Lorentz transformed tetrad ${\bf E}_{a'}$ will solve the equations too; only the global Lorentz invariance is guaranteed. This issue is harmless for the metric, whose invariance under local Lorentz transformations of the tetrad is not affected [see Eq.~\eqref{metric1} below]. Instead it means that $f(T)$ theories provide the field of tetrads with some degree(s) of freedom (d.o.f.) beyond those already contained in the metric. In fact, for a given metric, the theory selects a subset of tetrads among the entire set of tetrads matching the metric. The new d.o.f. should imply some property of gravity  not described by the metric. The physical nature of this property, and the way it couples to matter, are items that remain unsolved.

The issue of the d.o.f. of the theory has been addressed in some works, from the Hamiltonian formalism perspective \cite{Li:2011rn}, conformal transformations \cite{Yang:2010ji,Wright:2016ayu}, a remnant group of Lorentz transformations \cite{Ferraro:2014owa}, a null tetrad approach \cite{Bejarano:2014bca,Bejarano:2017akj}, among others. In \cite{Li:2011rn} it is established that $f(T)$ gravity has $\frac{n(n-3)}{2}+n-1$ d.o.f. in $n$ dimensions, and that the additional $n-1$ d.o.f. could be related to a massive vectorial field or a massless vector field plus a scalar field. However, there is no clue about the transformation that would manifest these objects. To the best of our knowledge, there is no work that indicates that $f(T)$ gravity has more than one additional degree of freedom, and therefore the claim made in \cite{Li:2011rn} should be revised in the light of new research. This is the main motivation of this work.

The outline of this paper is as follows. In Sec. \ref{sec:tegr-fT} we introduce the TEGR theoretical framework and $f(T)$ gravity. In Sec. \ref{sec:ham} we introduce the reader to the theory of constrained Hamiltonian systems and the Dirac-Bergmann algorithm. In Sec. \ref{sec:htegr} we review the Hamiltonian formulation of the teleparallel equivalent of general relativity. With this at hand, we perform the canonical formalism for $f(T)$ gravity in Sec. \ref{sec:hfT}, then in Sec. \ref{sec:dof} we study the consequences of the previous analysis in the counting of d.o.f. of the theory. Finally, we devote Sec. \ref{sec:concl} to the conclusions, and include an Appendix with a toy model that is useful to understand some features of the Hamiltonian analysis of $f(T)$ gravity.

\section{TEGR and $f(T)$}
\label{sec:tegr-fT}

\subsection{The teleparallel equivalent of general relativity}

We start by defining a manifold $M$, a basis $\{\mathbf{e}_{a}\}$ of vectors in the tangent space $T_{p}(M)$, and a dual basis $\{\mathbf{E}^{a}\}$ in the cotangent space $T^{*}_{p}(M)$. This means that the application of the one-forms $\mathbf{E}^{a}$ over the vectors $\mathbf{e}_{b}$ yields $\mathbf{E}^{a}(\mathbf{e}_{b})=\delta _{b}^{a}$. The vector basis can be expanded in a coordinate basis as $\mathbf{e}_{a}=e_{a}^{\mu }~\partial_{\mu }$ and $\mathbf{E}^{a}=E_{\mu }^{a}\ dx^{\mu }$. With this, the duality relation looks like
\begin{equation}
E_{\mu }^{a}\ e_{b}^{\ \mu }~=~\delta_{b}^{a}~,\ \ \ \ \ e_{a}^{\mu
}~E_{\nu }^{a}~=~\delta _{\nu }^{\mu }~.  \label{duality}
\end{equation}

Throughout this work we will denote spacetime coordinate indices by Greek letters $\mu ,\nu ,\ldots=0,\ldots,n-1$, Lorentzian tangent space indices by Latin letters $a,b,\ldots,g,h=0,\ldots,n-1$, and spatial coordinate indices by $i,j,\ldots = 1,\ldots, n-1$.

The metric field is expressed in terms of the \textit{vielbein} (vierbein o tetrad
in $n=4$ dimensions), which encodes the metric of the spacetime through the following relation
\begin{equation}
\mathbf{g}= \eta_{ab}\ \mathbf{E}^{a}\otimes \mathbf{E}^{b},
\label{metric1}
\end{equation}
then,
\begin{equation}
\mathbf{e}_{a}\cdot \mathbf{e}_{b} = \mathbf{g}(\mathbf{e}_{a},
\mathbf{e}_{b}) = \eta_{ab},
\end{equation}
which declares the vielbein an orthonormal basis. Written in component
notation, the previous expressions are
\begin{equation}
g_{\mu \nu }~=~\eta_{ab}~E_{\mu }^{a}~E_{\nu }^{b}~,\hspace{0.25in}\eta
_{ab}~=~g_{\mu \nu }~e_{a}^{\mu }~e_{b}^{\nu }~.
\end{equation}
In the same way, the relation between the metric volume and the determinant of
the matrix $E_{\mu }^{a}$ is
\begin{equation}
\sqrt{|g|}~=~\text{det}[E_{\mu }^{a}]~\doteq ~E~.  \label{volume}
\end{equation}

We formulate a dynamical theory for the spacetime by defining a Lagrangian that depends on the vielbein field. The Lagrangian leading to dynamical equations that are equivalent to Einstein equations is \cite{Hayashi:1979qx}
\begin{equation}
L = E\ T.  
\label{TEGR}
\end{equation}
This is the Lagrangian for the teleparallel equivalent of general relativity. Here $T$ is the torsion scalar
\begin{equation}
T~\doteq ~T_{\ ~\mu \nu }^{\rho }\ S_{\rho }^{\ ~\mu \nu },  \label{Tscalar}
\end{equation}
which is made up of the torsion of the Weitzenb\"{o}ck connection $\Gamma_{\ \nu \rho }^{\mu }\doteq e_{a}^{\ \mu }\,\partial
_{\nu }E_{\rho }^{a}$ \cite{Ortin2007},
\begin{equation}
T_{\hspace{0.05in}\,\nu \rho }^{\mu }~\doteq ~e_{a}^{\ \mu }~(\partial _{\nu
}E_{\rho }^{a}-\partial _{\rho }E_{\nu }^{a})~,  \label{TorsionTensor}
\end{equation}
and the so-called superpotential
\begin{equation}
S_{\rho }^{\ \mu \nu }~\doteq ~\dfrac{1}{2}\left( K_{\hspace{0.05in}\hspace{0.05in}\rho }^{\mu \nu }+T_{\lambda }^{\ \lambda \mu }~\delta _{\rho }^{\nu
}-T_{\lambda }^{\ \lambda \nu }~\delta _{\rho }^{\mu }\right) ~,
\end{equation}
where the contorsion associated with the Weitzenb\"{o}ck connection is
\begin{equation}
K_{\hspace{0.05in}\hspace{0.05in}\rho }^{\mu \nu }~\doteq ~\frac{1}{2} (T_{\rho }^{\hspace{0.05in}\mu \nu }-T_{\hspace{0.05in}\hspace{0.05in}\rho
}^{\mu \nu }+T_{\hspace{0.05in}\hspace{0.05in}\rho }^{\nu \mu }).
\label{contorsion}
\end{equation}

One of the most appealing features of the spacetime traced by the Weitzenb\"{o}ck connection is that it parallel-transports the vielbein along any curve, since $\nabla _{\nu }E_{\mu }^{a}=\partial _{\nu }E_{\mu }^{a}- \Gamma_{\ \nu \mu}^{\lambda } E_{\lambda }^{a}=0$. Besides, from Eq.\eqref{duality}, it is $\nabla_{\nu } e_{a}^{\mu }=0$. Therefore, parallelism can be thought as an absolute, path-independent notion that compares the projections $V^{a}=\mathbf{E}^{a}(\mathbf{V})$ of vectors on the vielbein. It is accomplished that $\nabla _{\nu }\mathbf{V}=\nabla _{\nu
}(V^{a}\mathbf{e}_{a})=\mathbf{e}_{a}~\partial _{\nu }V^{a}$, therefore the vector is parallel transported if and only if the components $V^a$ are constant. This property also implies that the Weitzenb\"{o}ck connection is metric compatible.

Another remarkable feature of the Weitzenb\"{o}ck connection is that it is curvatureless. The curvature two-form is given in terms of the spin connection $\omega^a_{\ b}$ as $\mathbf{R}_{\ b}^{a} \doteq d \omega_{\ b}^{a} + \omega_{\ c}^{a} \wedge \omega_{ \ b}^{c}$. The Weitzenb\"{o}ck connection is the choice $\omega^a_{\ b}=0$, which trivially vanishes $\mathbf{R}^a_{\ b}$. Notice that $\Gamma_{\ \nu \rho }^{\mu }$ does not vanish, since it is transformed to a coordinate basis. Instead, the curvature is a tensor, so it vanishes in any basis. Consequently, TEGR is a theory that encodes the gravitational effects in the torsion; in contrast GR describes a spacetime equipped with the torsionless Levi-Civita connection $\overline{\Gamma}_{\ \nu \rho }^{\mu }$ that has a nonvanishing curvature. While the Einstein-Hilbert Lagrangian is defined in terms of the Levi-Civita scalar curvature as $L=E~\overline{R}$, the TEGR Lagrangian depends on the torsion scalar $T$ that is built from the Weitzenb\"{o}ck connection. The relation between them is given by
\begin{equation}
-E\ \overline{R} = E\ T - 2\ \partial _{\rho }(E\ T_{\mu }^{\ \mu \rho }),
\label{RLC}
\end{equation}
which means that they only differ in a surface term. This is integrated out when put in the action, so guaranteeing the equivalence. \footnote{For the role of spacetime boundaries in teleparallel gravity, see \cite{Oshita:2017nhn}} However, the vielbein field has $n^2$ independent components, but the metric tensor has only $\frac{n(n+1)}{2}$. But TEGR dynamical equations are invariant under local Lorentz transformations of the vielbein. These transformations have $\binom{n}{2}$ generators, therefore this gauge invariance means that $\frac{n(n-1)}{2}$ d.o.f. are canceled out, thus agreeing with the fact that the theory is equivalent to GR at the level of the equations of motion.

\subsection{$f(T)$: modified teleparallel gravity}

Since the torsion scalar is quadratic in first-order derivatives of the vielbein field, any theory constructed with a function of it will lead to second-order dynamical equations for the vielbein. This is a suitable feature for a modified gravity model, and it was one of the motivations to propose the so-called $f(T)$ gravity or modified teleparallel gravity. The action that defines this theory is
\begin{equation}
\mathcal{S}[\mathbf{E}^a] = \dfrac{1}{2\kappa} \int d^4 x \ E \ f(T) + \int d^4 x \ E \ \mathcal{L}_{m},
\end{equation}
where the second term is a Lagrangian for matter. The dynamical equations of this theory are obtained varying this action with respect to the tetrad field, which are
\begin{equation}
\begin{split}
 & 4 e_a^{\ \lambda} S_{\lambda}^{\ \mu\nu} \partial_{\mu} T f''(T) - e_a^{\ \nu} f(T)\\
 & + 4 \left[e_a^{\ \lambda} T^{\rho}_{\ \mu\lambda} S_{\rho}^{\ \mu\nu} + e \partial_{\mu}(E e_a^{\ \lambda} S_{\lambda}^{\ \mu\nu} ) \right] f'(T)  = - 2 \kappa e_a^{\ \lambda} \mathcal{T}_{\lambda}^{\ \nu},
 \end{split}
 \label{eomft}
\end{equation}
where the matter energy-momentum tensor is $\mathcal{T}_{\lambda}^{\ \nu}$.
By virtue of the relation \eqref{RLC}, it is simple to see that any nonlinear function $f(T)$  will manifest nonlocal Lorentz invariance. This is because the four-divergence appearing in the equivalence \eqref{RLC} is not invariant, and will remain encapsulated in the functional form. This is an important issue and a common source of misunderstanding. Actually the lost of this local invariance means that by performing a local Lorentz transformation of a solution $\{ \mathbf{e}^a \}$ one will not necessarily obtain another solution, irrespective that the metric would not suffer any change. In other words, the dynamical equations \eqref{eomft} describe d.o.f. beyond the ones involved in the metric tensor. However the theory is still invariant under global Lorentz transformations.

\section{Constrained Hamiltonian systems}

\label{sec:ham}

We will briefly review Dirac's procedure for constrained Hamiltonian systems in field theory \cite{Anderson:1951ta,Dirac1964,Sundermeyer:1982gv,Sundermeyer:2014kha}. Given an action $S$ written in terms of a Lagrangian that depends on the canonical variables $L  = L (q^k, \dot{q}^k )$, the canonical momenta are defined as
\begin{equation}
 p_k ( q^k, \dot{q}^k ) = \dfrac{\partial L }{\partial \dot{q}^k}.
 \label{pmom}
\end{equation}
In a constrained physical system, not all the canonical momenta are linearly independent, but they will be related with the canonical coordinates through relations of the form
\begin{equation}
 \phi_{\rho}(q^k,p_k)=0, \ \ \ \ \rho=1,\ldots, P.
\end{equation}
Those are denominated primary constraints, and they appear at the level of the definitions \eqref{pmom}, and before using the equations of motion. They delimit a subset $\Gamma_c$ in the phase space of the theory, the so-called primary constraint surface.
The canonical Hamiltonian is defined in the standard way as
\begin{equation}
 H_c = \dot{q}^k p_k - L(q^k,\dot{q}^k),
\end{equation}
where a sum over $k$ is implicit. We also define the primary Hamiltonian
\begin{equation}
 H_p = H + u^{\rho} \phi_{\rho}.
\end{equation}
The $u^{\rho}$ are Lagrange multipliers; when varied independently, they ensure the primary constraints.

The preservation of the primary constraints over time is obtained through the primary Hamiltonian; this leads to the following equations:
\begin{equation}
 \begin{split}
  \dot{\phi}_{\rho'} & = \{ \phi_{\rho'}, H_p\} \\
  & = \{ \phi_{\rho'}, H_c \} + \{\phi_{\rho'},\phi_{\rho} \} u^{\rho} \overset{!}{\approx} 0.
 \end{split}
\end{equation}
If we define $h_{\rho'} \equiv \{ \phi_{\rho'}, H_p\}$ and $C_{\rho'\rho} \equiv \{\phi_{\rho'},\phi_{\rho} \}$, the solution of this system will depend on the values of these objects. In particular, new constraints will arise \footnote{Other cases are analyzed, for example, in \cite{Sundermeyer:1982gv,Sundermeyer:2014kha,Henneaux:1992ig}.} if $h_{\rho'} \not\approx 0$ and $\text{det}(C_{\rho'\rho}) \approx 0$. $C_{\rho'\rho}$ is a $P\times P$ matrix; if it has $\text{Rank}(C_{\rho'\rho})=M$, then it will be $P-M$ linearly independent null eigenvectors $\omega^{\rho'}_{\beta}$, with $\beta = 1,\ldots,P-M$, that impose
\begin{equation}
 \omega^{\rho'}_{\beta} \cdot h_{\rho'} \overset{!}{\approx} 0.
\end{equation}
These equations are either trivially satisfied or they originate new $S'$ constraints,
\begin{equation}
 \phi_{\overline{\rho}} \approx 0, \ \ \ \ \ \overline{\rho} = P+1, \ldots, P+S',
\end{equation}
which are called secondary constraints. This procedure must be iterated with the secondary constraints, which could originate new secondary constraints (sometimes called tertiary constraints), which repeat the procedure. The algorithm finishes when the following occurs: it exists a hypersurface $\Gamma_C$ in the phase space, defined by
\begin{equation}
\begin{split}
 \phi_{\rho} & \approx 0, \ \ \ \ \rho = 1, \ldots, P,\\
 \phi_{\overline{\rho}} & \approx 0, \ \ \ \ \overline{\rho} = P+1, \ldots,P+S.
 \end{split}
\end{equation}
The two sets contain all the $P$ primary constraints and all the $S$ secondary constraints (and tertiary, etc.). It is convenient to use a common notation for all of them, with $\phi_{\hat{\rho}}$, where $\hat{\rho}=1,\ldots,P+S$. With this, for each left null  eigenvector $\omega^{\hat{\rho}}_{\beta}$ of the matrix $C_{\hat{\rho}\rho} = \{ \phi_{\hat{\rho}}, \phi_{\rho} \}$, the following conditions are satisfied:
\begin{equation}
 \omega^{\hat{\rho}}_{\beta} \cdot \{ \phi_{\hat{\rho}}, H_c \} \approx \left|_{  \Gamma_C } 0 \right. .
\end{equation}
For the Lagrange multipliers, the following equations are fulfilled:
\begin{equation}
 \{ \phi_{\hat{\rho}}, H_c \} + \{ \phi_{\hat{\rho}}, \phi_{\rho} \} u^{\rho} \approx \left|_{\Gamma_C}  0 \right. .
 \label{usys}
\end{equation}
Some of these equations will be satisfied identically, others will represent conditions over the $u^{\rho}$'s. The case in which we are will depend on the rank of the matrix $C$. If $\text{Rank}(C_{\hat{\rho}\rho})=P$, then all Lagrange multipliers are fixed. If $\text{Rank}(C_{\hat{\rho}\rho})=K<P$, then it will be $P-K$ solutions to the equation
\begin{equation}
 C_{\hat{\rho}\rho} V^{\rho}_{\alpha} = \{ \phi_{\hat{\rho}}, \phi_{\rho} \} V^{\rho}_{\alpha} \approx 0, \ \ \ \ \ \alpha=1,\ldots,P-K.
 \label{homsys}
\end{equation}
The most general solution to the system \eqref{usys} is
\begin{equation}
 u^{\rho} = U^{\rho} + \mathcal{v}^{\alpha} V^{\rho}_{\alpha},
 \label{usol}
\end{equation}
where $U^{\rho}$ is a particular solution and $\mathcal{v}^{\alpha}$ are arbitrary coefficients that multiply the solutions of the homogeneous system \eqref{homsys}.

It is convenient to classify the constraints into \textit{first and second-class constraints}. It is said that a constraint is first-class if its Poisson bracket with all the constraints vanishes weakly. If a constraint is not first-class (there is at least one Poisson bracket that does not vanish), it is second-class. Any physical theory must be reformulated in terms of the maximum number of first-class constraints (and second-class constraints). We denote the set of first-class constraints as $\Phi_I, \ \ \ I = 1,\ldots, L$, and the remaining second-class constraints as $\chi_A$.

If we did this procedure correctly, then the matrix of Poisson brackets between second-class constraints, defined as
\begin{equation}
 \Delta_{AB} = \{ \chi_A, \chi_B \},
\end{equation}
should be invertible (i.e., its determinant should be different from zero). If not, then there is a first-class constraint hidden among the $\chi_A$ and it should be removed from the set by redefining the basis of constraints. Notice that the number of second-class constraints must be even, since otherwise $\text{det}(\Delta_{AB})=0$.
The total Hamiltonian $H_T$ is defined by using Eq. \eqref{usol}, so obtaining
\begin{equation}
 H_T = H' + \mathcal{v}^{\alpha} \phi_{\alpha},
\end{equation}
where $H' = H_c + U^{\rho} \phi_{\rho}$.
The system of equations \eqref{usys} is satisfied trivially for the first-class constraints, while for the second-class constraints it is written as
\begin{equation}
 \{ \chi_A, H_c \} + \Delta_{AB} u^B \approx 0.
\end{equation}
From this we can solve for the Lagrange multipliers, obtaining that
\begin{equation}
 u^B = \overline{\Delta}^{BA}\{\chi_A, H_c \},
\end{equation}
where $\overline{\Delta}^{BA}$ is the inverse of the matrix $\Delta_{AB}$. The result of this procedure is that all multipliers associated with a primary second-class constraint in $H'$ are determined, and the only free parameters are $\mathcal{v}^{\alpha}$. Therefore, there will be as many arbitrary functions in the Hamiltonian as first-class constraints exist. We are ready to calculate the number of d.o.f. in terms of the number of first and second-class constraints. It is that
\begin{equation}
\begin{split}
 \text{Number of d.o.f.}& = \text{Number of } (p,q) - \text{Number of f.c.c.}\\
 & - \dfrac{1}{2} \left( \text{Number of s.c.c.} \right).
 \end{split}
\end{equation}

\section{Review on the Hamiltonian formulation of TEGR}
\label{sec:htegr}

In this section we are going to summarize the main results of Ref. \cite{Ferraro:2016wht}. Here the TEGR Lagrangian was written in the form
\begin{equation}
\mathcal{L} = E\ T=\frac{1}{2} E\ \partial_{\mu } E_{\,\,\nu }^{a}\ \partial _{\rho
}E_{\,\,\lambda }^{b}\ e_{c}^{\mu } e_{e}^{\nu } e_{d}^{\rho
} e_{f}^{\lambda }\,M_{ab}^{\ \,\,\,\,cedf},
\label{LagrM}
\end{equation}
where the \emph{supermetric}, a Lorentz invariant tensor, was defined as
\begin{equation}
M_{ab}^{\ \,\,\,cedf} \doteq 2 \eta_{ab} \eta^{c[d} \eta^{f]e} - 4 \delta
_{a}^{[d} \eta^{f][c} \delta_{b}^{e]} + 8 \delta_{a}^{[c} \eta
^{e][d} \delta _{b}^{f]}.
\label{supermetric}
\end{equation}
The teleparallel equivalent of general relativity is, of course, a constrained system where not all the canonical momenta can be solved in terms of the velocities. The definition of the momenta, starting from the Lagrangian \eqref{LagrM}, can be written conveniently as
\begin{equation}
\Pi _{a}^{\mu }\,E_{\mu }^{e}\ =\ E\,C_{ab}^{\ \,\,\,\,ef}\,e_{f}^{\lambda
}\,\partial _{0}E_{\lambda }^{b}+E\ \partial _{i}E_{\lambda
}^{b}\,e_{c}^{0}\,e_{d}^{i}\,e_{f}^{\lambda }\,M_{ab}^{\ \,\,\,cedf},
\label{invariantmom}
\end{equation}
where the noninvertible $C_{ab}^{\ \,\,\,\,ef}$ matrix is defined as
\begin{equation}
C_{ab}^{\ \,\,\,\,ef}~\doteq ~e_{c}^{0}\ e_{d}^{0}\ M_{ab}^{\ \,\,\,cedf}~.
\label{C}
\end{equation}
From this definition we can obtain the following primary constraints:
\begin{equation}
\begin{split}
G^{(1)}_a & = \Pi^0_a \approx 0, \\
G^{(1)}_{ab} & = 2 \eta_{e[b} \Pi^i_{a]} E^e_i + 4 E \partial_i E^c_{j} e^0_{[b} e^i_a e^j_{c]} \approx 0.
\end{split}
\end{equation}
The first $n$ constraints indicate that the $E^a_0$ components of the vielbein are spurious gauge-dependent variables, which is in consonance with the fact that the temporal sector of the metric tensor has always a spurious character. The second $n(n-1)/2$ constraints represent the invariance of the theory under the Lorentz group in the tangent space. These constraints form a closed algebra that is precisely the Lorentz algebra.

After finding these constraints, it remains to find the time evolution. This is achieved by writing an expression for the primary Hamiltonian. For this, in \cite{Ferraro:2016wht} it proved to be useful the introduction of a new notation in which we arrange the components of $C_{ab}^{\ \,\,\,\,ef}$ in a $n^{2}\times n^{2}$ symmetric matrix $C_{AB}$, through the following identification
\begin{equation}
A=(a-1)\,n+e,\ \ \ \ B=(b-1)\,n+f,  \label{indexation}
\end{equation}
and then to compute the eigenvalues of $C_{\,\,B}^{A}$ in order to calculate the Moore-Penrose pseudoinverse $D^{AB}$. Here $C_{\,\,B}^{A} = \eta^{AC} C_{CB}$, where $\eta^{AC}=\eta^{ac} \eta_{eg}$ acts as a Minkowski metric in the space delimited by the multi-index notation. The eigenvalues of the matrix $C_{\,\,B}^{A}$ follow a very simple pattern: $n(n+1)/2$ eigenvalues are null, $n(n-1)/2-1$ of them are equal to $2\,g^{00}\doteq \lambda $, and the
remaining one is equal to $(2-n)\,\lambda $.

We build the canonical Hamiltonian by identifying the subset of canonical velocities that can be still solved in terms of the momenta. The momenta are written, by using the multi-index notation, as
\begin{equation}
 \Pi_A - P_A = E C_{AB} (\dot{E}^B-E^B_0).
 \label{pmomfull}
\end{equation}
Here it is defined
\begin{equation}
\begin{split}
&\dot{E}^B = e^{\lambda}_f \dot{E}^b_{\lambda}, \ \ \ \ E^B_0 = e^i_f \partial_i E^b_0, \\
&\Pi_A = \Pi^{\mu}_a E^e_{\mu}, \ \ \ \ P_A = E \partial_i E^b_k e^0_c e^i_d e^k_f M_{ab}^{\ \ cedf}.
\end{split}
\end{equation}
By using the pseudoinverse $D^{AB}$ to solve for the $\dot{E}^A$ in \eqref{pmomfull}, it is obtained that
\begin{equation}
 \dot{E}^A - E^A_0 = e D^{AB}(\Pi_B - P_B).
\end{equation}
Given that the Lagrangian is
\begin{equation}
 \mathcal{L} = \dfrac{1}{2}(\Pi_A + P_A)(\dot{E}^A - E^A_0) - U,
\end{equation}
where we also define
\begin{equation}
 U = - \dfrac{1}{2} E \partial_i E^a_j \partial_k E^b_l e^i_c e^j_e e^k_d e^l_f M_{ab}^{\ \ cedf},
\end{equation}
the canonical Hamiltonian density is written as
\begin{equation}
\begin{split}
 \mathcal{H} & = \Pi_A \dot{E}^A - \mathcal{L} \\
  & = \dfrac{1}{2} e (\Pi_A - P_A) D^{AB} (\Pi_B - P_B) + \Pi_A E^A_0 + U.
\end{split}
 \end{equation}
With this we can write the primary Hamiltonian, which would serve the purpose of evaluating the consistency of all the constraints over time. It is
\begin{equation}
 H_p = \int d^3 x\ \mathcal{H} + \int d \mathbf{x}\ u^a(t,\mathbf{x}) G^{(1)}_a(t,\mathbf{x}).
\end{equation}

The consistency of the whole procedure requires the following secondary constraints:
\begin{equation}
 G^{(2)}_0 = \mathcal{H} - \partial_i(E^c_0 \Pi^i_c) \approx 0,
 \label{G20}
\end{equation}
\begin{equation}
 G^{(2)}_k = \partial_k E^c_i \Pi^i_c - \partial_i(E^c_k \Pi^i_c) \approx 0.
\end{equation}
These are nothing but the super-Hamiltonian and supermomenta constraints of the Arnowitt-Deser-Misner (ADM) formulation of general relativity. It is worth noticing that while the ADM Hamiltonian vanishes on the constraint surface, the TEGR Hamiltonian does not. This is since the GR and TEGR Lagrangians differs by a surface term, then it follows in \eqref{G20} that $\mathcal{H}$ is not zero but a divergence.

The only nonvanishing Poisson brackets are the following. Firstly we have that
\begin{equation}
  \{G^{(2)}_0(t,\textbf{x}), G^{(1)}_a(t,\textbf{y}) \} =  (e^0_a G^{(2)}_0 + e^i_a G^{(2)}_i) \delta( \textbf{x} - \textbf{y}).
\end{equation}
The super-Hamiltonian and supermomenta constraints form the ADM algebra, given by
\begin{equation}
 \begin{split}
  \{G^{(2)}_i(t,\textbf{x}), G^{(2)}_j(t,\textbf{y}) \} & = - G^{(2)}_i(\textbf{x})\ \partial^{\textbf{y}}_j \delta(\textbf{x}-\textbf{y})  \\
  & + G^{(2)}_j(\textbf{y})\ \partial^{\textbf{x}}_i \delta(\textbf{x}-\textbf{y}), \\
  \{G^{(2)}_0(t,\textbf{x}), G^{(2)}_0(t,\textbf{y}) \} & = g^{ij}(\textbf{x}) G^{(2)}_i(\textbf{x})\ \partial^{\textbf{y}}_j \delta(\textbf{x}-\textbf{y}) \\
  & - g^{ij}(\textbf{y}) G^{(2)}_i(\textbf{y})\ \partial^{\textbf{x}}_j \delta(\textbf{x}-\textbf{y}), \\
  \{G^{(2)}_0(t,\textbf{x}), G^{(2)}_i(t,\textbf{y}) \} & = G^{(2)}_0(\textbf{x})\ \partial^{\textbf{y}}_i \delta(\textbf{x}-\textbf{y}). \\
 \end{split}
\end{equation}
Also, the generators of the Lorentz group form the Lorentz algebra
\begin{equation}
\begin{split}
 \{G^{(1)}_{ac}(t,\textbf{x}), G^{(1)}_{fe}(t,\textbf{y}) \} & = ( \eta_{ec} G^{(1)}_{af} + \eta_{af}G^{(1)}_{ce} \\
 & - \eta_{cf}G^{(1)}_{ae} - \eta_{ae} G^{(1)}_{cf} )\delta(\textbf{x}-\textbf{y}),
 \end{split}
\end{equation}
as expected. Finally, it is obtained
\begin{equation}
 \{ G^{(2)}_0(t,\textbf{x}), G^{(1)}_{ab}(t,\textbf{y}) \} = E^c_0 \eta_{c[a} e^0_{b]} G^{(2)}_0 \delta(\textbf{x}-\textbf{y}).
\end{equation}
The last result reflects the fact that the Hamiltonian, which is contained inside the constraint $G^{(2)}_0$, is not Lorentz invariant due to the four-divergence contained in the Lagrangian. Notice that the calculation of this bracket is new and did not appear on foregoing work, probably because of the different definitions of the constraints. However, we find our approach the simplest one, and therefore the best starting point for the Hamiltonian formulation of $f(T)$ gravity.

Finally we obtain that the entire set of constraints is first-class, and that they generate gauge transformations on the vielbein field. Moreover, there are $\frac{n(n+3)}{2}$ spurious variables, which reduces the number of d.o.f. to $\frac{n(n-3)}{2}$.

\section{Hamiltonian formulation of $f(T)$ gravity}
\label{sec:hfT}

\subsection{Degrees of freedom}

We will use the procedure developed in the preceding section as the base of the Hamiltonian formalism for $f(T)$ gravity, in order to perform the counting of d.o.f. of the theory, and with the aim of understanding its physical nature. It must be said that what they are and how these d.o.f. manifest themselves is an unsolved dilemma.  Previous work that attempted a Hamiltonian analysis of $f(T)$ gravity based on a different approach found that the theory had \textit{five} d.o.f. in four dimensions, i.e. three extra when compared with general relativity, or $\frac{n(n-3)}{2}+n-1$ d.o.f. in dimension $n$ \cite{Li:2011rn}. The authors suggested that these d.o.f. would manifest in a kind of Higgs mechanism, through a massive vectorial field or a scalar field plus a massless vectorial field. However, it has not been shown the equivalence between these fields and the $f(T)$ action. Moreover, no more than one extra degree of freedom appears at the level of cosmological perturbations \cite{Li:2011wu,Izumi:2012qj,Chen:2010va}.

There have been several proposals for understanding the issue of the d.o.f.. The recent finding of the remnant group of local Lorentz transformations in $f(T)$ gravity \cite{Ferraro:2014owa} makes a classification of pairs of tetrads and Lorentz matrices that satisfy the condition of preserving the otherwise Lorentz variating term of the $f(T)$ action. The null tetrad approach introduced in \cite{Bejarano:2014bca} in order to facilitate the search for GR geometries preserved in $f(T)$ has led to the finding of two tetrads that lead to the same Friedmann-Lemaître-Robertson-Walker metric, but with different torsion scalar \cite{Bejarano:2017akj}. This fact could be a manifestation of the extra d.o.f. of the theory, which remains to be understood.

A different approach is the \textit{covariant} $f(T)$ gravity \cite{Krssak:2015oua}, where a frame-dependent spin connection is chosen from that tetrad which is inertial when gravity is switched off. This procedure would render the torsion covariant under local Lorentz transformations of the tetrad; thus the local Lorentz invariance of the dynamics would be restored. There are other covariant approaches, as the Lagrange multiplier formulation suggested in \cite{Nester:2017wau}. However, there would be several ways of obtaining covariance in modified teleparallel gravities \cite{Golovnev:2017dox}. We do not know for sure if a covariant version of $f(T)$ gravity would manifest the additional d.o.f., as the noncovariance feature of the theory (i.e. the loss of local Lorentz invariance) is presumably the generator of these d.o.f., as we will see later through the constraint algebra of the theory. This is one of the reasons that discourage us from adopting a covariant approach in the first instance, but this issue should also be addressed by further research.

\subsection{Scalar equivalence}

It will be useful to rephrase $f(T)$ gravity as a scalar-tensor theory, by taking the following action containing the vierbein $E^a_{\mu}$ and an auxiliary scalar field $\phi$:
\begin{equation}
 \mathcal{S} = \dfrac{1}{2\kappa} \int d^4 x E [ \phi T- V(\phi) ] + \mathcal{S}_m(E^{a}_{\mu},\Psi ),
 \label{Leg}
\end{equation}
where $V(\phi)$ is a potential for the field $\phi$, and $\mathcal{S}_m(E^{a}_{\mu},\Psi)$ is the action for matter fields. By varying the action with respect to $\phi$ one obtains $T=V'(\phi)$, so linking the scalar field with the torsion scalar (and therefore the tetrad).
This relation between $T$ and $V(\phi)$ shows that the action \eqref{Leg} is dynamically equivalent to an action defined by the Lagrangian density $\mathcal{L} = E f(T)$. That is,
\begin{equation}
 \mathcal{L} = E(\phi T - V(\phi)) = E \left(\phi\dfrac{dV}{d\phi} - V(\phi) \right) = E f(T) 
\end{equation}
is the Legendre transform of the function $V(\phi)$; so, not only is it $T=V'(\phi)$ but $\phi=f'(T)$ as well. It can be seen that the limiting case of TEGR is obtained when $\phi=1$ and $V(\phi)=0$. The action \eqref{Leg} resembles the Jordan frame action in general relativity, but with the scalar $\phi$ accompanying the torsion scalar $T$ instead of the Ricci scalar. We will use the action \eqref{Leg} as a starting point for the Hamiltonian formalism for $f(T)$ gravity, keeping in mind that we just introduced an additional canonical coordinate $\phi$.

\subsection{Hamiltonian and primary constraints}

The canonical coordinates would be the set of $n^2+1$ functions $( \phi, E^a_{\mu} ) $. However, since the Lagrangian does not depend on the time derivatives of the variables $\phi$ or $E^a_0$, then the following primary constraints are obtained:
\begin{equation}
G^{(1)}_{\pi} = \pi \approx 0 , \ \ \ \ \  G^{(1)}_a = \Pi^0_a \approx 0.
\end{equation}
Removing the $\Pi^0_a$ from the definition of the $\Pi^{\mu}_a$, we get that the rest of the canonical momenta are given by
\begin{equation}
 \Pi^i_a = \dfrac{\partial \mathcal{L} }{\partial (\partial_0 E^a_i)} = \phi E \partial_{\rho} E^b_{\lambda} e^0_c e^i_e e^{\rho}_d e^{\lambda}_f M_{ab}^{\ \ cedf}.
 \label{pimom}
\end{equation}
With this, the Poisson bracket among two fields $A,B$ is defined as
\begin{equation}
\begin{split}
 \{ A, B \} & = \int d\mathbf{z} \left( \dfrac{\delta A}{\delta E^a_i} \dfrac{\delta B}{\delta \Pi^i_a} - \dfrac{\delta A}{\delta \Pi^i_a} \dfrac{\delta B}{\delta E^a_i} \right. \\
 & \left. + \dfrac{\delta A }{\delta \phi} \dfrac{\delta B}{\delta \pi} - \dfrac{\delta A}{\delta \pi} \dfrac{\delta B}{\delta \phi} \right).
 \end{split}
\end{equation}
Therefore, the fundamental Poisson brackets among the canonical variables are given by
\begin{equation}
 \begin{split}
  \{E^a_{\mu}(\mathbf{x}), \Pi^{\nu}_b(\mathbf{y}) \} & = \delta^a_b \delta^{\mu}_{\nu} \delta^{(n-1)}(\mathbf{x}-\mathbf{y}), \\
  \{ \phi(\mathbf{x}), \pi(\mathbf{y}) \} & = \delta^{(n-1)}(\mathbf{x}-\mathbf{y}),
  \label{fundamPB}
 \end{split}
\end{equation}
while the remaining brackets among canonical variables and momenta are zero.

With the help of the expression for the canonical momenta written in the multi-index notation,
\begin{equation}
 \Pi_A - \phi P_A = \phi E C_{AB} ( \dot{E}^B - E^B_0 ),
 \label{PiAm}
\end{equation}
we can find the remaining primary constraints, by considering the null eigenvectors for $C_{AB}$. The eigenvectors $v_{|gh|e}^{\ \ \ \ \ a} = 2\delta^a_{[g} \eta_{h]e}$ generate the following primary constraints:
\begin{equation}
 G^{(1)}_{ab} = 2\eta_{e[b}\Pi^i_{a]}E^e_i + 4 \phi E \partial_i E^c_j e^0_{[b} e^i_a e^j_{c]} \approx 0.
\end{equation}
These are $\frac{n(n-1)}{2}$ primary constraints that are slightly different from the ones appearing in TEGR, because of the presence of $\phi$ in the last term.

In conclusion, there are $\frac{n(n-1)}{2}+n+1$ primary constraints $(G^{(1)}_{ab},G^{(1)}_c,G^{(1)}_{\pi} )$ that define a hypersurface $\Gamma$ in the phase space of the theory. To build the primary Hamiltonian, we need the canonical Hamiltonian, for which we will use again the multi-index notation. The Lagrangian density is written as
\begin{equation}
 \mathcal{L} = \dfrac{1}{2} (\Pi_A + \phi P_A)(\dot{E}^A-E^A_0) - \phi\ U + E V(\phi).
\end{equation}
Solving for $\dot{E}^A$ from the expression \eqref{PiAm} through the use of the pseudoinverse $D^{AB}$, we find that the Lagrangian, in terms of the canonical momenta is
\begin{equation}
 \mathcal{L} = \dfrac{1}{2\phi} e ( \Pi_A \Pi_B - \phi^2 P_A P_B) D^{AB}  - \phi U + E V(\phi).
\end{equation}
The Hamiltonian density is defined in the traditional form as $ \mathcal{H} = \pi \dot{\phi} + \Pi^i_c \dot{E}^c_i - \mathcal{L}$. However, using again the definition \eqref{PiAm}, we get
\begin{equation}
\begin{split}
 \mathcal{H}  = & \dfrac{e}{2 \phi} (\Pi_A - \phi P_A)(\Pi_B - \phi P_B)D^{AB} \\
 & + \Pi_A E^A_0 + \phi U - E V(\phi).
\end{split}
 \end{equation}
Therefore, the primary Hamiltonian is
\begin{equation}
 \mathcal{H}_p = \mathcal{H}  + u^{ab} G^{(1)}_{ab} + u^a G^{(1)}_a  + u^{\pi} G^{(1)}_{\pi}.
\end{equation}
The notation adopted will be $u^{\rho} = (u^{ab}, u^c, u^{\pi})$ and in general, the Greek index $\rho$ will label primary constraints and its associated Lagrange multipliers.

At this point it is important to remark the following expression for the torsion scalar:
\begin{equation}
\begin{split}
 T & = \dfrac{1}{2}e^2 \left( \dfrac{1}{\phi}\Pi_A+P_A \right) D^{AB} \left( \dfrac{1}{\phi}\Pi_B-P_B \right) - e U \\
 & = \dfrac{1}{2}e^2 \left( \dfrac{1}{\phi^2}  \Pi_A \Pi_B  -  P_A P_B  \right)D^{AB} - e U .
\end{split}
\end{equation}

\subsection{Consistency of primary constraints}

We will study the consistency relations of the primary constraints with the tools provided by the Dirac-Bergmann algorithm introduced in Sec. \ref{sec:htegr}. We impose the time evolution of the primary constraints by means of the following relations:
\begin{equation}
 \{G^{(1)}_{\rho'}, \mathcal{H}  \}  + \{ G^{(1)}_{\rho'}, G^{(1)}_{\rho} \} u^{\rho}\approx 0,
 \label{cgral}
\end{equation}
where $G^{(1)}_{\rho'}$ denotes the set of primary constraints.

For the primary constraint $G^{(1)}_{\pi}$, the  relevant Poisson brackets are
\begin{equation}
\begin{split}
 \{G^{(1)}_{\pi}(\mathbf{x}), G^{(1)}_c(\mathbf{y}) \} & = 0, \\
 \{G^{(1)}_{\pi}(\mathbf{x}), G^{(1)}_{\pi}(\mathbf{y}) \} & = 0, \\
 \{G^{(1)}_{\pi}(\mathbf{x}), G^{(1)}_{ab}(\mathbf{y}) \} & = F_{ab} \delta^{(n-1)}(\mathbf{x}-\mathbf{y}), \\
 \{G^{(1)}_{\pi}(\mathbf{x}), \mathcal{H}(\mathbf{y}) \} & = -F_{\phi} \delta^{(n-1)}(\mathbf{x}-\mathbf{y}),
 \end{split}
 \label{consistg1pi}
\end{equation}
where it has been defined
\begin{equation}
 \begin{split}
  F_{\phi} & = \dfrac{e}{2}\left(\dfrac{1}{\phi^2} \Pi_A \Pi_B D^{AB} - P_A P_B D^{AB} \right) - U - E\ \dfrac{\partial V(\phi)}{\partial \phi} \\
  & = E \left( T - \dfrac{\partial V(\phi)}{\partial \phi} \right),
 \end{split}
\end{equation}
and also
\begin{equation}
 \begin{split}
  F_{ab} & = 4 E\ \partial_i E^c_j\ e^0_{[b} e^i_a e^j_{c]}\\
  & = \dfrac{4}{3}E\ (T_j(e^0_b e^j_a - e^j_b e^0_a) + e^i_b\ e^j_a\ T^0_{\ ij} )
 \end{split}
\end{equation}
with $T_j = T^i_{\ ij}$, $T^0_{\ ij} = e^0_c (\partial_i e^c_j - \partial_j e^c_i)$. Since $F_{ab}$ has $\frac{n(n-1)}{2}$ components, we will arrange them as a ``vector'' $F_{\tilde{a}}$ such that the indices are ordered in an increasing way
\begin{equation}
 F_{\tilde{a}} = (F_{01}, F_{02}, \ldots, F_{(n-2)\ (n-1)}) \equiv (F_1, F_2, \ldots, F_{\frac{n(n-1)}{2}} ) .
\end{equation}
We will use both notations indistinctly, according to the context. We notice in  \eqref{consistg1pi} that the only Poisson brackets different from zero are the ones that involve $G^{(1)}_{ab}$ and $\mathcal{H}$, since they contain a functional dependence on $\phi$.

Next we compute the Poisson brackets with the primary constraint $G^{(1)}_c$, which are simpler:
\begin{equation}
 \begin{split}
& \{ G^{(1)}_c(t,\mathbf{x}), G^{(1)}_d(t,\mathbf{y}) \} = 0, \\
& \{ G^{(1)}_c(t,\mathbf{x}), G^{(1)}_{ab}(t,\mathbf{y}) \} = 0, \\
& \{ G^{(1)}_c(t,\mathbf{x}), \mathcal{H}(t,\mathbf{y}) \} = - (e^0_c G^{(2)}_0 + e^i_a G^{(2)}_i)\delta^{(n-1)}(\mathbf{x}-\mathbf{y}).
\end{split}
\end{equation}
We notice the appearance of two expressions $G^{(2)}_i$ and $G^{(2)}_0$ given by
\begin{equation}
 \begin{split}
  G^{(2)}_0 & = \mathcal{H} - \partial_i( E^c_0 \Pi^i_c ) \approx 0,\\
  G^{(2)}_k & = \partial_k E^c_i \Pi^i_c - \partial_i( E^c_k \Pi^i_c ) \approx 0,
 \end{split}
\end{equation}
which will be proven to be secondary constraints later. It remains to calculate the pertinent Poisson brackets for the $G^{(1)}_{ab}$, which resume to
\begin{equation}
 \begin{split}
 \{ G^{(1)}_{ab}(t,\mathbf{x}), G^{(1)}_{ef}(t,\mathbf{y}) \} & = (\eta_{eb} G^{(1)}_{af} + \eta_{af} G^{(1)}_{be} \\
 & - \eta_{ae} G^{(1)}_{bf} - \eta_{bf} G^{(1)}_{ae})\delta^{(3)}(\mathbf{x}-\mathbf{y}), \\
 \{ G^{(1)}_{ab}(t,\mathbf{x}), \mathcal{H}(t,\mathbf{y}) \} & = E^e_0 \eta_{e[b} e^0_{a]} G^{(2)}_0.
 \end{split}
\end{equation}
We see that, even though the Lorentz constraints are modified, they still satisfy the Lorentz algebra. Besides, we see that the expression $G^{(2)}_0$ takes part also in the second bracket.

The set of equations that would determine the Lagrange multipliers is the following:
\begin{equation}
 \begin{split}
\dot{G}^{(1)}_{c} & = - (e^0_c G^{(2)}_0 + e^i_a G^{(2)}_i) \approx 0, \\
\dot{G}^{(1)}_{ab} & = E^c_0 \eta_{c[b}e^0_{a]} G^{(2)}_0 + u^{fe}( \eta_{eb} G^{(1)}_{af} + \eta_{af} G^{(1)}_{be}\\
& - \eta_{bf} G^{(1)}_{ae}-\eta_{ae} G^{(1)}_{bf} ) + u^{\pi} F_{ab} \approx 0,  \\
\dot{G}^{(1)}_{\pi} & = F_{\phi} - u^{ab} F_{ab} \approx 0 .
\end{split}
\label{condprim}
\end{equation}
The general procedure for the Dirac-Bergmann algorithm requires that we write these as the following matricial system:
\begin{equation}
h_{\rho'} +  C_{\rho' \rho} u^{\rho} \approx 0,
\end{equation}
where $C_{\rho' \rho} = \{ \phi_{\rho'}, \phi_{\rho} \}$ is the matrix containing the Poisson brackets already calculated, and $h_{\rho'} = \{\phi_{\rho'}, \mathcal{H} \}$ is a vector containing the Poisson brackets between the primary constraints and the canonical Hamiltonian. This matrix is
\begin{equation}
 \begin{array}{c@{\hspace{-5pt}}l}
C_{\rho' \rho} = \left(
\begin{array}{c|c|c}
0 \ \ \ \ \cdots \ \ \ \ 0 \ & \ 0 \ \cdots \ 0 \ & \ -F_{1} \\
\ \ \ddots \ \ & \ \ \ddots \ \ & \vdots  \\
0 \ \ \ \ \cdots \ \ \ \ 0 \ & \ 0 \ \cdots \ 0 \ & \ -F_{\frac{n(n-1)}{2}} \\ \hline
0 \ \ \ \ \cdots \ \ \ \ 0 \ & \ 0 \ \cdots \ 0 \ & \ 0 \\
\ \ \ddots \ \ & \ \ \ddots \ \ &  \ \vdots \\
0 \ \ \ \ \cdots \ \ \ \ 0 \ & \ 0 \ \cdots \ 0 \ & \ 0 \\
\hline
F_{1} \cdots F_{\frac{n(n-1)}{2}} & \ 0 \ \cdots \ 0 \ & \ 0
\end{array}
\right)
&\begin{array}{l}\left. \rule{0mm}{7mm}\right\}{ \text{\scriptsize $\frac{n(n-1)}{2}$} }\\
\\ \left. \rule{0mm}{7mm}\right\}{ \text{\scriptsize $n$} } \vspace{6mm}
\end{array}\\[-5pt]
\begin{array}{cc}\underbrace{\rule{19mm}{0mm}}_{\frac{n(n-1)}{2}} \ & \ \underbrace{\rule{12mm}{0mm}}_{n}\hspace{20pt}
\end{array}&
\end{array}.
\label{Cgeneral}
\end{equation}
Besides, the vector $h_{\rho'}$ can be decomposed into three pieces: $\frac{n(n-1)}{2}$ components denoting the expression $\{G^{(1)}_{ab}, \mathcal{H} \} = e_{0[b} e^0_{a]} G^{(2)}_0$, $n$ components that represent the bracket  $\{G^{(1)}_c, \mathcal{H} \} = -e^0_c G^{(2)}_0 - e^i_c G^{(2)}_i$, and the last vectorial component is $\{ G^{(1)}_{\pi},\mathcal{H} \} = F_{\phi}$. Therefore, $h_{\rho'}$ is given by
\begin{equation}
\begin{split}
&h_{\rho'}  = (e_{0[0} e^0_{1]} G^{(2)}_0, \ldots,e_{0[0} e^0_{(n-1)]} G^{(2)}_0 ,e_{0[1} e^0_{2]} G^{(2)}_0,\ldots , \\
&e_{0[1} e^0_{(n-1)]} G^{(2)}_0,\ldots ,e_{0[(n-2)} e^0_{(n-1)]} G^{(2)}_0,-e^{\mu}_0 G^{(2)}_{\mu},\ldots ,\\
& -e^{\mu}_{n-1} G^{(2)}_{\mu} , F_{\phi} ).
\end{split}
\label{hrho}
\end{equation}
The algorithm requires to compute the left and right null eigenvectors of the matrix $C_{\rho'\rho}$. In this case, since this is a square antisymmetric matrix, these null eigenvectors coincide. It is easy to see that there are $n$ null eigenvectors $V^{\rho}_{\alpha}$ given by
\begin{equation}
 \begin{split}
  V^{\rho}_{\alpha=1} & = ( \underbrace{0, \ldots, 0}_{\frac{n(n-1)}{2}} , \underbrace{ 1 , 0, \ldots, 0}_{n} , 0), \\
  & \hspace{2mm} \vdots \\
  V^{\rho}_{\alpha=n} & = ( \underbrace{0, \ldots, 0}_{\frac{n(n-1)}{2}} ,  \underbrace{ 0, \ldots, 0, 1 }_{n} , 0).
  \label{auvecn}
 \end{split}
\end{equation}
There is still an additional set of null eigenvectors obtained through the following constraint on the components of $V^{\rho}_{\alpha} = (V^1,\ldots,V^{\frac{n(n-1)}{2} })$:
\begin{equation}
 F_1 \cdot V^1 + F_2 \cdot V^2 + \cdots + F_{\frac{n(n-1)}{2} } \cdot V^{\frac{n(n-1)}{2} } = 0,
 \label{cndFV}
\end{equation}
which is obtained from the last row of $C_{\rho'\rho}$. Since the rank of the matrix $C_{\rho'\rho}$ is always $2$, and we have already obtained $n$ null eigenvectors, we should be able to obtain $\frac{n(n-1)}{2}-1$ null eigenvectors from the condition \eqref{cndFV}. The choice of the components is completely arbitrary, as long as they are linearly independent to the set \eqref{auvecn} and constrained to satisfy \eqref{cndFV}. A possible choice is the following:
\begin{equation}
 \begin{split}
   V^{\rho}_{\alpha = n+1} & = ( \underbrace{ F_{2}, -F_{1}, 0, \ldots, 0}_{\frac{n(n-1)}{2}}, \underbrace{ 0, \ldots, 0 }_{n}, 0), \\
  & \hspace{2mm} \vdots \\
  V^{\rho}_{\alpha = n+\frac{n(n-1)}{2}-1} & = ( \underbrace{F_{\frac{n(n-1)}{2} }, 0, \ldots, 0, -F_{1} }_{\frac{n(n-1)}{2}}, \underbrace{ 0, \ldots, 0 }_{n}, 0).
 \end{split}
 \label{auvecl}
\end{equation}
No matter how we make this choice, the eigenvectors $V^{\rho}_{\alpha}$ always impose that $G^{(2)}_0 \approx 0$, since the following condition,
\begin{equation}
 V^{\rho}_{\alpha} \cdot h_{\rho'} \overset{!}{\approx} 0
\end{equation}
must be satisfied for all values of $\alpha$ [notice that in \eqref{hrho}, $h_{\rho'}$ is proportional to $G^{(2)}_0$ in the first $\frac{n(n-1)}{2}$ entries]. If $G^{(2)}_0$ is weakly zero, and we put this in the first equation in \eqref{condprim}, it is immediately obtained that also $G^{(2)}_i \approx 0$. Therefore, there are $n$ new secondary constraints $G^{(2)}_0, G^{(2)}_i$, whose consistency relations must be analyzed.

\subsection{Secondary constraints and consistency}

In order to impose the consistency of the secondary constraints correctly, we need to take into account the following Poisson brackets between primary and secondary constraints:
\begin{equation}
 \begin{split}
  \{ G^{(2)}_i(t,\mathbf{x}), G^{(1)}_{\pi}(t,\mathbf{y})\} &= 0, \\
  \{ G^{(2)}_i(t,\mathbf{x}), G^{(1)}_{c}(t,\mathbf{y}) \} & = 0,\\
  \{ G^{(2)}_i(t,\mathbf{x}), G^{(1)}_{ab}(t,\mathbf{y}) \} & = 0, \\
  \{ G^{(2)}_i(t,\mathbf{x}), G^{(2)}_0(t,\mathbf{y}) \} & = -G^{(2)}_0(\mathbf{y})\ \partial^{\mathbf{x}}_i\delta(\mathbf{x}-\mathbf{y}),\\
  \{ G^{(2)}_0(t,\mathbf{x}), G^{(2)}_0(t,\mathbf{y}) \} & = g^{ij}(\mathbf{x})G^{(2)}_i(\mathbf{x})\ \partial^{\mathbf{y}}_j\delta(\mathbf{x}\\
  & -\mathbf{y}) - g^{ij}(\mathbf{y})G^{(2)}_i(\mathbf{y})\ \partial^{\mathbf{x}}_j\delta(\mathbf{x}-\mathbf{y}),\\
  \{ G^{(2)}_i(t,\mathbf{x}), G^{(2)}_j(t,\mathbf{y}) \} & = -G^{(2)}_i(\mathbf{x})\ \partial^{\mathbf{y}}_j\delta(\mathbf{x}-\mathbf{y})\\
  & +G^{(2)}_j(\mathbf{y})\ \partial^{\mathbf{x}}_i\delta(\mathbf{x}-\mathbf{y}).
 \end{split}
\end{equation}
Then, the requirement of consistency over time of the secondary constraints can be written as the following system:
\begin{equation}
 \begin{split}
  \dot{G}^{(2)}_0 & = g^{ij} G^{(2)}_i(\mathbf{x})\  \partial^{\mathbf{y}}_j \delta(\mathbf{x}-\mathbf{y}) - g^{ij} G^{(2)}_i (\mathbf{y})\ \partial^{\mathbf{x}}_j \delta(\mathbf{x}-\mathbf{y}) \\
  &+ u^{ab} E^c_0 \eta_{c[b} e^0_{a]} G^{(2)}_0 + u^a(e^0_a G^{(2)}_0 + e^i_a G^{(2)}_i ) \\
  & + u^{\pi} F_{\phi} \approx 0, \\
  \dot{G}^{(2)}_i & = -G^{(2)}_0 \partial_i \delta(\mathbf{x}-\mathbf{y}) \approx 0.
 \end{split}
\end{equation}
These equations, on the new constraint surface, form the following conditions:
\begin{equation}
 \begin{split}
\dot{G}^{(1)}_{c} & \approx 0, \\
\dot{G}^{(1)}_{ab} & = u^{\pi} F_{ab} \approx 0,  \\
\dot{G}^{(1)}_{\pi} & = F_{\phi} - u^{ab} F_{ab} \approx 0, \\
\dot{G}^{(2)}_0 & = u^{\pi} F_{\phi} \approx 0, \\
\dot{G}^{(2)}_i & \approx 0.
\label{sistufinal}
 \end{split}
\end{equation}
This can be understood as an extended matricial system, where
\begin{equation}
 \begin{array}{c@{\hspace{-5pt}}l}
C_{\hat{\rho} \rho} \approx \left(
\begin{array}{c|c|c}
0 \ \ \ \ \cdots \ \ \ \ 0 \ & \ 0 \ \cdots \ 0 \ & \ -F_{1} \\
\ \ \ddots \ \ & \ \ \ddots \ \ & \vdots  \\
0 \ \ \ \ \cdots \ \ \ \ 0 \ & \ 0 \ \cdots \ 0 \ & \ -F_{\frac{n(n-1)}{2}} \\ \hline
0 \ \ \ \ \cdots \ \ \ \ 0 \ & \ 0 \ \cdots \ 0 \ & \ 0 \\
\ \ \ddots \ \ & \ \ \ddots \ \ &  \ \vdots \\
0 \ \ \ \ \cdots \ \ \ \ 0 \ & \ 0 \ \cdots \ 0 \ & \ 0 \\
\hline
F_{1} \cdots F_{\frac{n(n-1)}{2}} & \ 0 \ \cdots \ 0 \ & \ 0 \\
\hline
0 \ \ \ \ \cdots \ \ \ \ 0 \ & \ 0 \ \cdots \ 0 \ & \ F_{\phi} \\
\hline
0 \ \ \ \ \cdots \ \ \ \ 0 \ & \ 0 \ \cdots \ 0 \ & \ 0 \\
\ \ \ddots \ \ & \ \ \ddots \ \ &  \ \vdots \\
0 \ \ \ \ \cdots \ \ \ \ 0 \ & \ 0 \ \cdots \ 0 \ & \ 0 \\
\end{array}
\right)
&\begin{array}{l}
\vspace{2.5mm}
\left. \rule{0mm}{9mm}\right\}{ \text{\scriptsize $\frac{n(n-1)}{2}$} }\\
\vspace{1.5mm}
\left. \rule{0mm}{7.5mm}\right\}{ \text{\scriptsize $n$} } \\
\vspace{1mm}
\left. \rule{0mm}{2mm} \right\}{ \text{ \scriptsize $1$ }} \\
\vspace{1mm}
\left. \rule{0mm}{2mm} \right\}{ \text{ \scriptsize $1$ }} \\
\vspace{1mm}
\left. \rule{0mm}{8mm} \right\}{ \text{ \scriptsize $n-1$ }} \\
\end{array}\\[-5pt]
\begin{array}{cc}\underbrace{\rule{20mm}{0mm}}_{\frac{n(n-1)}{2}} \ & \ \underbrace{\rule{12mm}{0mm}}_{n}\hspace{20pt}
\end{array}&
\end{array},
\label{Cgeneralfinal}
\end{equation}
and
\begin{equation}
h_{\hat{\rho}} = (0,0,0,0,0,0,0,0,0,0,F_{\phi},0,0,0,0) .
\end{equation}
The null eigenvectors of the augmented matrix $C_{\hat{\rho}\rho}$ have a  different number of components depending on if they are right or left null eigenvectors. We denote the right null eigenvectors by $V^{\rho}_{\alpha}$ with $\frac{n(n-1)}{2}+n+1$ components, and the left null eigenvectors by $\omega^{\hat{\rho}}_{\beta}$ with $\frac{n(n-1)}{2}+2n+1$ components.
It is not hard to see that the right null eigenvectors of \eqref{Cgeneralfinal} are the same as the right null eigenvectors of \eqref{Cgeneral}. That is, now the $V^{\rho}_{\alpha}$ of\eqref{Cgeneralfinal} are given by \eqref{auvecn} and \eqref{auvecl}. On the other hand, the first $2n-1$ left null eigenvectors are given by
\begin{equation}
 \begin{split}
  \omega^{\rho'}_{\beta=1} & = ( \underbrace{0, \ldots, 0}_{\frac{n(n-1)}{2}} , \underbrace{ 1 , 0, \ldots, 0}_{n}, 0, 0, \underbrace{0, \ldots, 0}_{n-2}, 0), \\
  & \vdots \\
  \omega^{\rho'}_{\beta=n} & = ( \underbrace{0, \ldots, 0}_{\frac{n(n-1)}{2}} , \underbrace{ 0, \ldots, 0, 1 }_{n}, 0, 0, \underbrace{0, \ldots, 0}_{n-2}, 0),
  \end{split}
\end{equation}
\begin{equation}
 \begin{split}
    \omega^{\rho'}_{\beta=n+1} & = ( \underbrace{0, \ldots, 0}_{\frac{n(n-1)}{2}} , \underbrace{ 0, \ldots, 0}_{n}, 0, 0, \underbrace{1, 0, \ldots, 0}_{n-2}, 0), \\
  & \vdots \\
    \omega^{\rho'}_{\beta=2n-1}& = ( \underbrace{0, \ldots, 0}_{\frac{n(n-1)}{2}} , \underbrace{ 0, \ldots, 0}_{n}, 0, 0, \underbrace{0, \ldots, 0, 0}_{n-2}, 1).
 \end{split}
\end{equation}
Furthermore, there is a condition that generates $\frac{n(n-1)}{2}$ extra null eigenvectors, which is
\begin{equation}
-F_{1} \cdot \omega^1 - \cdots - F_{ \frac{n(n-1)}{2} } \cdot \omega^{\frac{n(n-1)}{2}} + F_{\phi} \cdot \omega^{\frac{n(n-1)}{2}+n+2 } = 0.
\label{wftcond2}
\end{equation}
The additional null eigenvectors can be selected in the following way:
\begin{small}
\begin{equation}
 \begin{split}
\omega^{\hat{\rho}}_{\beta = 2n} & = ( \underbrace{ F_{2}, -F_{1}, 0, \ldots, 0}_{\frac{n(n-1)}{2}}, \underbrace{ 0, \ldots, 0 }_{2n} , 0), \\
& \vdots \\
\omega^{\hat{\rho}}_{\beta = 2n+\frac{n(n-1)}{2}-2} & = ( \underbrace{F_{\frac{n(n-1)}{2}-1 }, 0, \ldots, -F_{1}}_{\frac{n(n-1)}{2}},  \underbrace{ 0, \ldots, 0 }_{2n} , 0), \\
\omega^{\hat{\rho}}_{\beta = 2n+\frac{n(n-1)}{2}-1} & = ( \underbrace{0, \ldots, -F_{\phi} }_{\frac{n(n-1)}{2}},  \underbrace{ 0, \ldots}_{n}, 0, F_{\frac{n(n-1)}{2}}, \underbrace{ 0, \ldots, 0 }_{n-2} , 0),
 \end{split}
\end{equation}
\end{small}
although this is one of many possible choices. A specific choice will not interfere with the Hamiltonian formalism, as long as the selected basis satisfies \eqref{wftcond2}.

The left null eigenvectors impose the conditions $\omega^{\hat{\rho}}_{\beta}~\cdot~h_{\hat{\rho}}~\overset{!}{\approx}~0$, however since the component $\omega^{\frac{n(n-1)}{2}+n+1 }$ is restricted to be zero, and the component $h_{\frac{n(n-1)}{2}+n+1}$ is precisely the only one that is different from zero, then the $\omega^{\hat{\rho}}_{\beta}$'s do not generate any new secondary (tertiary) constraint, and the algorithm is finished.

\subsection{First- and second-class constraints}

It remains to find the Lagrange multipliers, a problem that is linked with the separation between first- and second-class constraints, for if a multiplier is not determined by the equations of motion, it would be linked to a first-class constraint, and vice versa. The solution for the Lagrange multipliers can be written in the following way:
\begin{equation}
 u^{\rho} = U^{\rho} + v^{\rho} = U^{\rho} + \mathcal{v}^{\alpha} V^{\rho}_{\alpha},
\end{equation}
where $U^{\rho}$ stands for the particular solution to the system and $\mathcal{v}^{\alpha}$ are arbitrary coefficients, one for each null eigenvector. We denote by $v^{\rho}=\mathcal{v}^{\alpha} V^{\rho}_{\alpha}$ the solution to the homogeneous system. The right null eigenvectors $V^{\rho}_{\alpha}$ are given by \eqref{auvecn} and \eqref{auvecl}, which determine the following:
\begin{itemize}
 \item The set of $n$ right null eigenvectors $V^{\rho}_{\alpha}$, $\alpha=1,\ldots,n$, determines that the $u^c$'s associated to $G^{(1)}_c$ are not fixed and then generate gauge transformations.
 \item None of the $n+\frac{n(n-1)}{2}-1$ right null eigenvectors has the last component nonvanishing, henceforth it imposes $u^{\pi}=0$, as expected.
 \item The $\frac{n(n-1)}{2}$ remaining eigenvectors give the following relations among the $v^{ab}$ part of the  multipliers:
  \begin{equation}
  \begin{split}
   v^{01} & = \mathcal{v}_{2} F_2 + \mathcal{v}_{3} F_3 + \cdots + \mathcal{v}_{\frac{n(n-1)}{2}} F_{\frac{n(n-1)}{2}}, \\
   v^{02} & = - \mathcal{v}_{2} F_1, \\
   v^{03} & = - \mathcal{v}_{3} F_1, \\
   & \vdots \\
   v^{(n-2)\ (n-1)} & = - \mathcal{v}_{\frac{n(n-1)}{2}} F_1.
  \end{split}
  \label{solfinu}
 \end{equation}
\end{itemize}
This set of equations can be combined in a unique equation that relates all the multipliers, namely,
\begin{equation}
v^{01} F_1  + v^{02} F_2 + \cdots + v^{(n-2)\ (n-1)} F_{\frac{n(n-1)}{2}} = 0.
 \label{allues}
\end{equation}
This equation determines one of the $\frac{n(n-1)}{2}$ Lagrange multipliers, however we still have another relation to fulfill, which is
\begin{equation}
 F_{\phi} - U^{ab} F_{ab} \approx 0.
 \label{ufphi}
\end{equation}
The expression \eqref{ufphi} determines a hypersurface on which the $U^{ab}$ would be restricted, therefore there is one of the $U^{ab}$ that is fully determined. In order to illustrate this fact more clearly, and because any $U^{ab}$ can be chosen to satisfy \eqref{ufphi} without modifying the Hamiltonian formulation, we can work with
\begin{equation}
 U^{01} F_{1} = F_{\phi},
\end{equation}
and $U^{02},  \ldots, U^{(n-2)\ (n-1)} = 0$. Therefore, $u^{\pi}$ and $u^{01}$ would be the only multipliers that are determined through the procedure, which suggests that there are only two second-class constraints that remain to be found. This can be achieved by rewriting linear combinations of the primary and secondary constraints, so as to define combinations that commute with the rest of the constraints. Any choice of the form
\begin{equation}
 \tilde{G}^{(1)}_{ab} = \alpha^{cd}_{\ \ ab} G^{(1)}_{cd}
\end{equation}
will continue to be in the constraint surface, as the brackets $\{G^{(1)}_{ef}, \alpha^{cd}_{\ \ ab}\} G^{(1)}_{cd}$ are still multiplied by a primary constraint.  Using this argument, we can recombine the primary constraints associated to the Lorentz algebra as the following set:
\begin{equation}
 \begin{split}
 & \tilde{G}^{(1)}_{02} = F_{01} G^{(1)}_{02} - F_{02} G^{(1)}_{01}, \\
 & \tilde{G}^{(1)}_{03} = F_{02} G^{(1)}_{03} - F_{03} G^{(1)}_{01}, \\
 & \hspace{8mm} \vdots \\
 & \tilde{G}^{(1)}_{(n-2)\ (n-1)} = F_{01} G^{(1)}_{(n-2)\ (n-1)} - F_{(n-2)\ (n-1)} G^{(1)}_{01},
 \end{split}
 \label{fcclorentz}
\end{equation}
while $G^{(1)}_{01}$ remains unchanged. Any Poisson bracket of an element of this set with $G^{(1)}_{\pi}$ will vanish, except for the specific Lorentz constraint $G^{(1)}_{01} $, which would be the second-class constraint. However there is still another combination to be performed, since $\{ G^{(2)}_0, G^{(1)}_{\pi} \} = F_{\phi}$ would mean that either $G^{(2)}_0$ or $G^{(1)}_{\pi}$ are second-class. For this, we perform the following redefinition
\begin{equation}
 \begin{split}
&  \tilde{G}^{(2)}_0 = F_{01} G^{(2)}_0 - F_{\phi} G^{(1)}_{01}, \\
& \tilde{G}^{(1)}_{01} =  G^{(1)}_{01}.
 \end{split}
 \label{Gfirstclass}
\end{equation}
From this we have found the linear combination of constraints $ \tilde{G}^{(2)}_0$ that render a first-class constraint, since
\begin{equation}
 \{  \tilde{G}^{(2)}_0, G^{(1)}_{\pi} \} = \dfrac{\partial F_{\phi}}{\partial\phi} G^{(1)}_{01} \approx 0,
\end{equation}
while the constraint $\tilde{G}^{(1)}_{01}$ is still second-class and linearly independent from $ \tilde{G}^{(2)}_0$. Notice that all the Poisson brackets of $ \tilde{G}^{(2)}_0 $ with the remaining constraints vanish on the constraint surface.

In this way, the matrix of constraints is zero by blocks, except for the block that contains the second-class constraints that we will denote $\Delta_{AB}$, while the second-class constraints themselves are denoted by $\chi_A = (G^{(1)}_{\pi}, \tilde{G}^{(1)}_{01} )$. The aforementioned matrix is
\begin{equation}
 \Delta_{AB} = \left(
 \begin{array}{cc}
  \{G^{(1)}_{\pi}, G^{(1)}_{\pi} \} & \{ \tilde{G}^{(1)}_{01} , G^{(1)}_{\pi} \} \\
  \{G^{(1)}_{\pi}, \tilde{G}^{(1)}_{01} \} & \{ \tilde{G}^{(1)}_{01}, \tilde{G}^{(1)}_{01} \}
 \end{array}
 \right) = \left(
 \begin{array}{cc}
  0 & F_{01} \\
  - F_{01} & 0
 \end{array}
 \right).
\end{equation}
The system of equations $h_{\hat{\rho}} + C_{\hat{\rho}\rho} u^{\rho}$ is a trivial identity for the first-class constraints, while for the $\chi_A$'s this system can be read
\begin{equation}
 \{ \chi_A, H_c \} + \Delta_{AB} u^{B} \overset{!}{\approx} 0.
\end{equation}
From this we can solve for the Lagrange multipliers $u^B~=~(u^{\pi},~u^{01})$ associated to the $\chi_A$, since the equations
\begin{equation}
\begin{split}
& \{ G^{(1)}_{\pi}, H_c \} + \Delta_{\pi B} u^B = -F_{\phi} + F_{01} u^{01} \approx 0, \\
& \{ G^{(1)}_{01}, H_c \} + \Delta_{A\ (01)} u^A =  - F_{01} u^{\pi} \approx 0
\end{split}
\end{equation}
impose that $u^{\pi} \approx 0$ and $ F_{01} u^{01} \approx F_{\phi}$.

\section{Degrees of freedom of $f(T)$ gravity}
\label{sec:dof}

Finally we are able to determine the number of d.o.f. of the theory in dimension $n$. Notice that our formalism was always dimension independent. The counting of d.o.f. goes in the following way. We have
\begin{itemize}
 \item $n$ first-class constraints $G^{(1)}_c$,
 \item $n-1$ first-class constraints $G^{(2)}_i$,
 \item $\frac{n(n-1)}{2}-1$ first-class constraints $\tilde{G}^{(1)}_{ab}$ defined in \eqref{fcclorentz},
 \item $1$ first-class constraint $\tilde{G}^{(2)}_0$,
 \item $2$ second-class constraint $G^{(1)}_{\pi}$ and $\tilde{G}^{(1)}_{01}$.
\end{itemize}
Therefore, we have that
\begin{equation}
\begin{split}
 \text{d.o.f.} & =\#\, \text{pairs of canonical variables}  \notag \\
& - \#\,\text{first-class constraints}  \notag \\
& -\frac{1}{2}\ \#\,\text{second-class constraints}\ \\
& = n^2 + 1 - \left( 2n-1+\dfrac{n(n-1)}{2} - 1+1\right)-1 \\
& = \dfrac{n(n-3)}{2}+1.
\end{split}
\label{dof}
\end{equation}
Therefore, regardless of the number of dimensions, $f(T)$ gravity has only one extra degree of freedom in comparison with TEGR (or GR). Here we obtained that the local Lorentz invariance is lost in only one generator of Lorentz transformations, \textit{which is not specified} by the Hamiltonian formalism. This generator could be a combination of boosts and rotations that are fixed by the theory. It is noticeable that in order to preserve the first-class definition of the super-Hamiltonian constraint, it is required to redefine it by making use of the same Lorentz constraint that is fixed by the theory [see Eq. \eqref{Gfirstclass}]. This can also be understood as a mechanism for the breakdown of the Lorentz invariance, however in order to fully understand this matter, we require additional research to be performed. As far as we know this result could be considered a forerunner in the field of modified teleparallel gravity, and the techniques developed in \cite{Ferraro:2016wht} and in this work should be extended for wider classes of gravity theories with teleparallel structure, for example teleparallel gravities presented in \cite{Ferraro:2008ey,Gonzalez:2015sha,Bahamonde:2017wwk,Hohmann:2018rwf,Nester:2017wau} among others.

\subsection{Discussion of previous works}

Regarding the work of Li, Miao and Miao \cite{Li:2011rn}, we obtain a different number of d.o.f.. There are substantial differences between their work and ours, first the fact that the Hamiltonian formulation of TEGR from which we started is different. Their work is based in a first-order formulation developed in \cite{Maluf:1998ae,Maluf:2001rg,daRochaNeto:2011ir}, and therefore their primary and secondary constraints are different from ours, although we have almost the same number of constraints. The main difference lies in their secondary constraint $\pi_1 = \text{det}(M) \approx 0$, which does not appear in our formalism.
The square matrix $M$ in their work contains the Poisson brackets between their constraints $H_0$, $\Gamma^{ab}$ and $\pi$; these constraints are their analog to our constraints $G^{(2)}_0$, $G^{(1)}_{ab}$ and $G^{(1)}_{\pi}$, respectively. The reason why we believe that $\pi_1$ is not a secondary constraint is the following: the authors in \cite{Li:2011rn} define a system of equations $M\Lambda = 0$ for the Lagrange multipliers $\Lambda$, and they assert that the system must have a solution for \textit{all} the multipliers. The condition for the system to have a solution is $\text{det}(M)=0$, then they affirm that this should be a secondary constraint $\pi_1\approx 0$ (a very complicated one). We believe that this secondary constraint should not exist since, as we have proven in our formalism, not all the Lagrange multipliers are determined and moreover, there are only two that are determined through the procedure. Their attempt in order to calculate the Poisson brackets with this constraint lead the authors to assume that $g_{\mu\nu}$ is diagonal. Their claim is that this condition is not a gauge, but a technique of calculation. However this means an additional restriction over the vielbein field, which is a truly secondary constraint, and should have been taken into account into their counting of d.o.f.

Another discussion on the issue of the d.o.f. of $f(T)$ gravity can be found in \cite{Chen:2014qtl}, even though they do not perform any calculation on the Poisson brackets of the theory.  The authors argue different possibilities for the number of d.o.f., based in a hand-waving Hamiltonian analysis. They speculate what would be if a certain number of the Lorentz constraints became second-class. Considering that in $n=4$ there are eight constraints associated to the super-Hamiltonian, supermomenta, and the $\Pi^0_a\approx 0$ constraints, the authors imagine three possible cases, under the assumption that there are not additional constraints and taking pairs of Lorentz constraints to become second-class. The possible outcomes are $5$, $4$ and $2$ d.o.f. for $f(T)$ gravity, if six, four and two Lorentz constraints become second-class, respectively. Nonetheless, our work shows that under a mathematical equivalence the theory is interpreted as having an extra canonical variable that generates an extra primary constraint, which turns out to be second-class and turn one Lorentz constraint to be second-class. This possibility was not contemplated by \cite{Chen:2014qtl} although the constraint $G^{(1)}_{\pi} = \pi \approx 0$ was already considered in \cite{Li:2011rn}.

\section{Conclusions}
\label{sec:concl}

To develop the Hamiltonian formalism for $f(T)$ gravity, we have started from a scalar-tensor action that is mathematically equivalent to the action for $f(T)$ gravity. This involves a canonical variable $\phi$ that generates an additional primary constraint $G^{(1)}_{\pi}$. This constraint alters the constraint structure by making one of the $\frac{n(n-1)}{2}$ Lorentz constraints $G^{(1)}_{ab}$ to become second-class; however which constraint $G^{(1)}_{ab}$ becomes second-class is not \textit{a priori} determined, and any linear combination of them could play that role. The super-Hamiltonian constraint needs to be redefined using this Lorentz constraint in order to preserve its first-class attribute. All constraints proved to be first-class except for one of the Lorentz constraints and $G^{(1)}_{\pi}$. Since there are $n^2+1$ pairs of canonical variables, $\frac{n(n-1)}{2}+2n-1$ primary constraints and two  secondary constraints in any dimension, it results that $f(T)$ gravity has $\frac{n(n-3)}{2}+1$ d.o.f. This is $1$  extra degree of freedom when compared with TEGR gravity, which could be related to a scalar field. This might mean that the Einstein frame for $f(T)$ gravity could exist, but only for certain conditions that have not been established, an issue that will be investigated in future work \cite{Ferraro2018b}.

Although both $f(T)$ and metric $f(R)$ theories have an only additional degree of freedom, the way this extra degree of freedom appears is different in each case. In $f(R)$ gravity, it comes from the fact that the dynamical equations are fourth order; usually one lows the order by introducing the scalar field $\phi=f'(R)$ to cast the equations to a second order system with an additional degree of freedom $\phi$. Instead $f(T)$ gravity has dynamical equations of second order. However, the teleparallel theories come with an extended algebra of constraints; besides the generators of diffeomorphisms that are shared with $f(R)$ and other theories of gravity, the algebra includes the generators of the Lorentz group. This is because the tetrad involves more components than the metric. Thus the number of extra components that are pure gauge depends on how many Lorentz constraints are first-class. In TEGR all the Lorentz constraints are first-class, so reducing the number of d.o.f. to those contained in the metric. In $f(T)$ theories a linear combination of Lorentz constraints becomes second-class, thereby increasing by one the number of physical d.o.f.

\section*{ACKNOWLEDGMENTS}

The authors thank C. Bejarano, F. Fiorini, M. Kr\v{s}\v{s}\'ak and M. Lagos for helpful discussions. This work was supported by Consejo Nacional de Investigaciones Cient\'ificas y T\'ecnicas (CONICET) and Universidad de Buenos Aires. R. F. is a member of Carrera del Investigador Cient\'{\i}fico (CONICET, Argentina).

\bigskip \appendix

\section{A toy model: a pseudoinvariant rotational theory}
\label{sec:toymodel}
\subsection{Pseudoinvariant rotational Lagrangian}

Some features of the Hamiltonian formalism for $f(T)$ gravity can be observed in a simpler physical model: a Lagrangian that possesses rotational pseudoinvariance. Let us consider the following Lagrangian
\begin{equation}
 L = A \dfrac{\dot{z}}{z} + B \dfrac{\dot{\overline{z}}}{\overline{z}} + U(z\overline{z}),
 \label{Lrot}
\end{equation}
where $z=x+iy$ and $\overline{z}$ are the canonical variables, $A$ and $B$ are constants, and $U(z\overline{z})$ is a potential. This Lagrangian has rotational pseudoinvariance, since under a local rotation $e^{i\alpha(t)}$, the quotient $\dot{z}/z$ transforms as
\begin{equation}
 \dfrac{\dot{z}}{z} \longrightarrow \dfrac{\dot{z}}{z} + i \dot{\alpha}.
\end{equation}
In fact, after this transformation, the Lagrangian acquires a boundary term whenever $A$ is different from $B$.

The canonical momenta for the Lagrangian \eqref{Lrot} are
\begin{equation}
 \dfrac{\partial L}{\partial \dot{z}} \equiv p_z = \dfrac{A}{z}, \ \ \ \ \ \dfrac{\partial L}{\partial \dot{\overline{z}}} \equiv p_{\overline{z}} = \dfrac{B}{\overline{z}},
\end{equation}
which define two primary constraints given by
\begin{equation}
 G^{(1)}_{z} \equiv p_z - \dfrac{A}{z} \approx 0, \ \ \ \ \ G^{(1)}_{\overline{z}} \equiv p_{\overline{z}} - \dfrac{B}{\overline{z}} \approx 0.
\end{equation}
It is simple to see that the canonical Hamiltonian will be given by $H=U(z\overline{z})$, thus the primary Hamiltonian is
\begin{equation}
 H_p = U(z \overline{z}) + u^z \left( p_z - \dfrac{A}{z} \right) + u^{\overline{z}} \left( p_{\overline{z}} - \dfrac{B}{\overline{z}} \right),
\end{equation}
where $u^z$ and $u^{\overline{z}}$ are the Lagrange multipliers associated with $G^{(1)}_z$ and $G^{(1)}_{\overline{z}}$, respectively. We test the consistency over time of the primary constraints by calculating the Poisson bracket of them with $H_p$, and impose the result to be zero. Then, one obtains
\begin{equation}
 \begin{split}
  \{ G^{(1)}_{z} , H_p \} & = -U^{\prime} \overline{z} \overset{!}{\approx} 0 \equiv G^{(2)}_{z}, \\
  \{ G^{(1)}_{\overline{z}} , H_p \} & = -U^{\prime} z \overset{!}{\approx} 0 \equiv G^{(2)}_{\overline{z}},
 \end{split}
\end{equation}
where it has been defined $U'=\frac{\partial U}{\partial(z\overline{z} )}$. We are in a case of reducible constraints, that is we can write a relation of dependence among two or more constraints. In this particular case it follows that
\begin{equation}
 z G^{(2)}_{z} = \overline{z} G^{(2)}_{\overline{z}} \equiv G^{(2)},
\end{equation}
where $G^{(2)}$ will be the independent secondary constraint that will be  taken into account in the formalism. Next we calculate its time evolution, which is given by
\begin{equation}
\begin{split}
 \dot{G}^{(2)}  = \{ G^{(2)}, H_p \} = & u^z (\overline{z} U^{\prime} + z \overline{z}^2 U^{\prime\prime}) \\
 & + u^{\overline{z}}( z U^{\prime} + z^2 \overline{z} U^{\prime\prime}) \approx 0,
 \end{split}
\end{equation}
which constrains the Lagrange multipliers to satisfy the dependence  relation $u^z = -\frac{z}{\overline{z}} u^{\overline{z}}$. One Lagrange multiplier is undetermined, therefore one of the primary constraints should be first-class. This can be better understood if we choose another basis for the subspace of primary constraints. In particular, we define the following:
\begin{equation}
 \begin{split}
G^{(1)}_a & \equiv \dfrac{1}{2} \left( z G^{(1)}_{z} - \overline{z} G^{(1)}_{\overline{z}} \right) = \dfrac{1}{2}\left(z p_z - \overline{z} p_{\overline{z}} + (B-A) \right), \\
G^{(1)}_b & \equiv \dfrac{1}{2} \left( z G^{(1)}_{z} + \overline{z} G^{(1)}_{\overline{z}} \right) = \dfrac{1}{2}\left( z p_z + \overline{z} p_{\overline{z}} - (B+A) \right).
\end{split}
\end{equation}
In this new basis, $G^{(1)}_a$ commutes with the other constraints, while the Poisson bracket
\begin{equation}
 \{ G^{(1)}_b, G^{(2)} \} = - z^2 \overline{z}^2 U^{\prime\prime}
\end{equation}
states that both constraints are second-class. Therefore, the  $2$ d.o.f. spanned by $(z,\overline{z})$ are removed, and the theory has no dynamics and it is pure gauge. The primary Hamiltonian in the new basis is
\begin{equation}
\begin{split}
 H_p = & U(z\overline{z}) + \dfrac{u^a}{2} \left( z p_z - \overline{z} p_{\overline{z}} + (B-A) \right) \\
 & + \dfrac{u^b}{2} \left( z p_z + \overline{z} p_{\overline{z}} - (B+A) \right),
 \end{split}
\end{equation}
and the consistency relation for $G^{(2)}$
\begin{equation}
 \begin{split}
  \dot{G}^{(2)} & = - u^b \left( z^2 \overline{z}^2 U^{\prime\prime} \right) \overset{!}{\approx} 0
 \end{split}
\end{equation}
imposes that $u^b=0$, while $u^a$ is indetermined and generates a gauge transformation given by $G^{(1)}_a$.

\subsection{Modified pseudoinvariant rotational Lagrangian}

Now we take a Lagrangian that is a function of the pseudoinvariant Lagrangian, i.e. $L=f(A\frac{\dot{z}}{z} + B \frac{\dot{\overline{z}}}{\overline{z}} + U(z\overline{z}))$.  This new theory can be worked out with the help of a scalar field $\phi$ such that
\begin{equation}
 L = \phi \left(A\frac{\dot{z}}{z} + B\frac{\dot{\overline{z}}}{\overline{z}} + U(z\overline{z}) \right) - V(\phi).
\end{equation}
The equation of motion for $\phi$ gives $A\frac{\dot{z}}{z} + B \frac{\dot{\overline{z}}}{\overline{z}} + U(z\overline{z}) - V^{\prime}(\phi) = 0$.

We write the canonical momenta of the theory, and obtain three primary constraints given by
\begin{equation}
 \begin{split}
p_z & = \dfrac{A\phi}{z} \  \longrightarrow \ G^{(1)}_z \equiv p_z - \dfrac{A\phi}{z} \approx 0, \\
p_{\overline{z}} & = \dfrac{B\phi}{\overline{z}} \  \longrightarrow \ G^{(1)}_{\overline{z}} \equiv p_{\overline{z}} - \dfrac{B\phi}{\overline{z}} \approx 0, \\
\pi & = 0 \equiv G^{(1)}_{\pi}.
 \end{split}
\end{equation}
Then, the primary Hamiltonian is
\begin{equation}
\begin{split}
 H_p =& \phi\ U(z\overline{z})\ +\ V(\phi)\ +\ u^z\left( p_z - \dfrac{A\phi}{z} \right)\\
 & + u^{\overline{z}}\left( p_{\overline{z}} - \dfrac{B\phi}{\overline{z}} \right) + u^{\pi} \pi.
 \end{split}
\end{equation}
With this, we write the consistency relations for the three primary constraints, which result:
\begin{equation}
\begin{split}
 \dot{G}^{(1)}_{z} & = -\phi \overline{z} U^{\prime}  - u^{\pi} \dfrac{A}{z} \approx 0, \\
 \dot{G}^{(1)}_{\overline{z}} & = -\phi z U^{\prime}  - u^{\pi} \dfrac{B}{\overline{z} } \approx 0, \\
 \dot{G}^{(1)}_{\pi} & = -U - \dfrac{dV}{d\phi} + u^z \dfrac{A}{z} + u^{\overline{z}} \dfrac{B}{\overline{z} } \approx 0.
 \end{split}
 \end{equation}
The formal procedure, where one must find the null eigenvector of the matrix of constraints, is applicable here. For this, we write this system in matricial form as
\begin{equation}
\begin{split}
&  h_{\rho'} + u^{\rho} C_{\rho'\rho}  \approx 0,  \\
& \left(
\begin{array}{c}
-\phi \overline{z} U^{\prime} \\
-\phi z U^{\prime} \\
-U - \frac{dV}{d\phi}
\end{array}
\right)
+
\left(
 \begin{array}{ccc}
  0 & 0 & -A/z \\
  0 & 0 & -B/\overline{z} \\
  A/z & B/\overline{z} & 0
 \end{array}
\right)
\left(
\begin{array}{c}
u^{z} \\
u^{\overline{z}} \\
u^{\pi}
\end{array}
\right) \approx 0.
\end{split}
\end{equation}
The left null eigenvectors of the matrix $C_{\rho'\rho}$ will determine conditions over the $u^{\rho}$, or give rise to new constraints. In this case there is only one (left and right) null eigenvector $V^{\rho'}_{(1)} = (zB, -\overline{z}A,0)$, which imposes the condition
\begin{equation}
 V^{\rho'} \cdot h_{\rho'}  = -\phi z \overline{z} U^{\prime} (B-A) \overset{!}{\approx} 0 \equiv - G^{(2)},
\end{equation}
which is a genuine secondary constraint. This constraint appears in the case $B \neq A$, otherwise the Lagrangian possesses total invariance under rotations, a case that we set aside. Therefore, we define $G^{(2)} \equiv \phi z \overline{z} U^{\prime}$, and study its consistency through the equation
\begin{equation}
\begin{split}
 \dot{G}^{(2)} & = \{G^{(2)},H_p \}  = u^z \phi \overline{z}( U^{\prime} + z \overline{z} U^{\prime\prime}) \\
 & + u^{\overline{z}} \phi z ( U^{\prime} + z \overline{z} U^{\prime\prime} ) + u^{\pi} z \overline{z} U^{\prime} \overset{!}{\approx} 0.
 \end{split}
\end{equation}
When we add this new relation of consistency to the matrix $C_{\rho'\rho}$, it does not produces new secondary constraints. Therefore, the system is reduced to
\begin{equation}
\begin{split}
 \dot{G}^{(1)}_{z} & =  - u^{\pi} \dfrac{A}{z} \approx 0, \\
 \dot{G}^{(1)}_{\overline{z}} & =  - u^{\pi} \dfrac{B}{\overline{z} } \approx 0, \\
 \dot{G}^{(1)}_{\pi} & = -U - \dfrac{dV}{d\phi} + u^z \dfrac{A}{z} + u^{\overline{z}} \dfrac{B}{\overline{z} } \approx 0, \\
 \dot{G}^{(2)} & = u^z \phi \overline{z}(U^{\prime} + z \overline{z} U^{\prime\prime}) + u^{\overline{z}} \phi z ( U^{\prime} + z \overline{z} U^{\prime\prime} )\\
 & + u^{\pi} z \overline{z} U^{\prime} \overset{!}{\approx} 0 .
 \end{split}
 \label{cmodrot}
 \end{equation}
From this we conclude that all constraints are second-class and remove $2$ d.o.f., leaving the theory with only~$1$~true degree of freedom. The following values for the Lagrange multipliers,
\begin{equation}
\begin{split}
 u^{\pi} & = 0, \\
 u^z & = -\dfrac{z}{B-A}\left( U + \dfrac{dV}{d\phi} \right), \\
 u^{\overline{z}} &= \dfrac{\overline{z}}{B-A}\left( U + \dfrac{dV}{d\phi} \right),
\end{split}
\end{equation}
solve the system \eqref{cmodrot}. 
 \vspace{45mm}

\begingroup\raggedright\endgroup


\begin{thebibliography}{55}
\expandafter\ifx\csname natexlab\endcsname\relax\def\natexlab#1{#1}\fi

\bibitem[Ferraro and Fiorini(2007)]{Ferraro:2006jd}
R.~Ferraro and F.~Fiorini, Modified teleparallel gravity: Inflation without inflaton, {\em Phys. Rev.} {\bfseries D 75},  084031 (2007) \href{http://xxx.lanl.gov/abs/gr-qc/0610067}{{\ttfamily arXiv:gr-qc/0610067}}.

\bibitem[Weyl(1918)]{Weyl1918}
H.~Weyl, \textit{Das Relativitätsprinzip, Fortschritte der Mathematischen Wissenschaften in Monographien} (Vieweg+Teubner Verlag, Berlin, 1918), pp. 147--159.

\bibitem[Einstein(1925)]{Einstein1925}
A.~Einstein, Einheitliche Feldtheorie der Gravitation und Elektrizität,  Pruess. Akad. Wiss. {\bfseries 414} (1925).

\bibitem[Einstein(1928)]{Einstein1928}
A.~Einstein, {\em Sitzungsberichte der Preussischen Akademie der Wissenschaften} (Verlag der Akademie der Wissenschaften, Berlin, 1928), pp. 217--221.

\bibitem[Einstein({1930})]{Einstein1930a}
A.~Einstein, Zur Theorie der Räume mit Riemannmetrik und Fernparallelismus,  Sitzungsber. Preuss. Akad. Wiss. Phys. Math. Kl. {\bfseries
  401} (1930).

\bibitem[Einstein(1930)]{Einstein1930b}
A.~Einstein, Auf die Riemann-Metrik und den Fern-Parallelismus gegründete einheitliche Feldtheorie, Math. Ann. {\bfseries 102}, 685 (1930).

\bibitem[Cartan(1922{\natexlab{a}})]{Cartan1922a}
E.~Cartan, Sur une généralisation de la notion de courbure de Riemann et les espaces à torsion, C.R. Hebd. Seances Acad. Sci. {\bfseries 174}, 593 (1922).

\bibitem[Cartan(1922{\natexlab{b}})]{Cartan1922b}
E.~Cartan, Sur les espaces généralisés et la théorie de la Relativité, C.R. Hebd. Seances Acad. Sci. {\bfseries 174}, 734 (1922).

\bibitem[Weitzenböck(1923)]{Weitzenbock1923}
R.~Weitzenböck, \textit{Invarianten theorie} (Noordhoff, Groningen, 1923).

\bibitem[Cho(1976)]{Cho:1975dh}
Y.~M. Cho, Einstein Lagrangian as the translational Yang-Mills
  Lagrangian, Phys. Rev. D {\bfseries 14}, 2521 (1976).

\bibitem[Linder(2010)]{Linder:2010py}
E.~V. Linder, Einstein's Other Gravity and the Acceleration of the
  Universe, Phys. Rev. D {\bfseries 81}, 127301 (2010),
  \href{http://xxx.lanl.gov/abs/1005.3039}{{\ttfamily arXiv:1005.3039}},
[Erratum: Phys. Rev.D 82, 109902 (2010)].

\bibitem[Bamba et~al.(2010)Bamba, Geng, and Lee]{Bamba:2010iw}
K.~Bamba, C.-Q. Geng, and C.-C. Lee, Comment on `Einstein's other gravity
  and the acceleration of the Universe',
 \href{http://xxx.lanl.gov/abs/1008.4036}{{\ttfamily arXiv:1008.4036}}.

\bibitem[Wu and Yu(2010)]{Wu:2010xk}
P.~Wu and H.~W. Yu, The dynamical behavior of $f(T)$ theory, Phys.
  Lett. B {\bfseries 692}, 176--179 (2010) 
 \href{http://xxx.lanl.gov/abs/1007.2348}{{\ttfamily arXiv:1007.2348}}.

\bibitem[Bamba et~al.(2011)Bamba, Geng, Lee, and Luo]{Bamba:2010wb}
K.~Bamba, C.-Q. Geng, C.-C. Lee, and L.-W. Luo, Equation of state for dark
  energy in $f(T)$ gravity, J. Cosmol. Astropart. Phys. 01 (2011) 021,
 \href{http://xxx.lanl.gov/abs/1011.0508}{{\ttfamily arXiv:1011.0508}}.

\bibitem[Yang(2011)]{Yang:2010ji}
R.-J. Yang, Conformal transformation in $f(T)$ theories, Europhys. Lett. 
  {\bfseries 93}, 60001 (2011) 
 \href{http://xxx.lanl.gov/abs/1010.1376}{{\ttfamily arXiv:1010.1376}}.

\bibitem[Li et~al.(2011)Li, Sotiriou, and Barrow]{Li:2010cg}
B.~Li, T.~P. Sotiriou, and J.~D. Barrow, $f(T)$ gravity and local Lorentz
  invariance, Phys. Rev. D {\bfseries 83}, 064035 (2011) 
 \href{http://xxx.lanl.gov/abs/1010.1041}{{\ttfamily arXiv:1010.1041}}.

\bibitem[Chen et~al.(2011)Chen, Dent, Dutta, and Saridakis]{Chen:2010va}
S.-H. Chen, J.~B. Dent, S.~Dutta, and E.~N. Saridakis, Cosmological
  perturbations in $f(T)$ gravity, Phys. Rev. D {\bfseries 83}, 023508 (2011)
 \href{http://xxx.lanl.gov/abs/1008.1250}{{\ttfamily arXiv:1008.1250}}.

\bibitem[Sotiriou et~al.(2011)Sotiriou, Li, and Barrow]{Sotiriou:2010mv}
T.~P. Sotiriou, B.~Li, and J.~D. Barrow, Generalizations of teleparallel
  gravity and local Lorentz symmetry, Phys. Rev. D {\bfseries 83}, 104030 (2011) \href{http://xxx.lanl.gov/abs/1012.4039}{{\ttfamily arXiv:1012.4039}}.

\bibitem[Li et~al.(2011{\natexlab{a}})Li, Sotiriou, and Barrow]{Li:2011wu}
B.~Li, T.~P. Sotiriou, and J.~D. Barrow, Large-scale Structure in $f(T)$
  Gravity, Phys. Rev. D {\bfseries 83}, 104017 (2011) \href{http://xxx.lanl.gov/abs/1103.2786}{{\ttfamily arXiv:1103.2786}}.

\bibitem[Li et~al.(2011{\natexlab{b}})Li, Miao, and Miao]{Li:2011rn}
M.~Li, R.-X. Miao, and Y.-G. Miao, Degrees of freedom of $f(T)$ gravity, J. High Energy Phys. {\bfseries 07} (2011) 108  \href{http://xxx.lanl.gov/abs/1105.5934}{{\ttfamily arXiv:1105.5934}}.

\bibitem[Izumi and Ong(2013)]{Izumi:2012qj}
K.~Izumi and Y.~C. Ong, Cosmological Perturbation in $f(T)$ Gravity
  Revisited, J. Cosmol. Astropart. Phys. {\bfseries 06} (2013) 029
 \href{http://xxx.lanl.gov/abs/1212.5774}{{\ttfamily arXiv:1212.5774}}.

\bibitem[Fiorini et~al.(2014)Fiorini, Gonz\'alez, and
  V\'asquez]{Fiorini:2013hva}
F.~Fiorini, P.~A. Gonz\'alez, and Y.~V\'asquez, Compact extra dimensions in
  cosmologies with $f(T)$ structure, Phys. Rev. D {\bfseries 89}, 024028
  (2014) \href{http://xxx.lanl.gov/abs/1304.1912}{{\ttfamily arXiv:1304.1912}}.

\bibitem[Fiorini(2013)]{Fiorini:2013kba}
F.~Fiorini, Nonsingular Promises from Born-Infeld Gravity, Phys.
  Rev. Lett. {\bfseries 111}, 041104 (2013) \href{http://xxx.lanl.gov/abs/1306.4392}{{\ttfamily arXiv:1306.4392}}.

\bibitem[Chen et~al.(2015)Chen, Izumi, Nester, and Ong]{Chen:2014qtl}
P.~Chen, K.~Izumi, J.~M. Nester, and Y.~C. Ong, Remnant Symmetry,
  Propagation and Evolution in $f(T)$ Gravity, Phys. Rev. D {\bfseries
   91}, 064003 (2015) 
 \href{http://xxx.lanl.gov/abs/1412.8383}{{\ttfamily arXiv:1412.8383}}.

\bibitem[Fiorini(2016)]{Fiorini:2015hob}
F.~Fiorini, Primordial brusque bounce in Born-Infeld determinantal
  gravity, Phys. Rev. D {\bfseries 94}, 024030 (2016) 
 \href{http://xxx.lanl.gov/abs/1511.03227}{{\ttfamily arXiv:1511.03227}}.

\bibitem[Gonz\'alez and V\'asquez(2015)]{Gonzalez:2015sha}
P.~A. Gonz\'alez and Y.~V\'asquez, Teleparallel Equivalent of Lovelock
  Gravity, Phys. Rev. D {\bfseries 92}, 124023 (2015) 
 \href{http://xxx.lanl.gov/abs/1508.01174}{{\ttfamily arXiv:1508.01174}}.

\bibitem[Cai et~al.(2016)Cai, Capozziello, De~Laurentis, and
  Saridakis]{Cai:2015emx}
Y.-F. Cai, S.~Capozziello, M.~De~Laurentis, and E.~N. Saridakis, $f(T)$
  teleparallel gravity and cosmology, Rept. Prog. Phys. {\bfseries 79}, 106901
   (2016) \href{http://xxx.lanl.gov/abs/1511.07586}{{\ttfamily arXiv:1511.07586}}.

\bibitem[Fiorini and Vattuone(2016)]{Fiorini:2016zrt}
F.~Fiorini and N.~Vattuone, An analysis of Born–Infeld determinantal
  gravity in Weitzenböck spacetime, Phys. Lett. B {\bfseries 763}, 45 
  (2016)  \href{http://xxx.lanl.gov/abs/1608.02622}{{\ttfamily arXiv:1608.02622}}.

\bibitem[Bahamonde and Böhmer(2016)]{Bahamonde:2016kba}
S.~Bahamonde and C.~G. Böhmer, Modified teleparallel theories of gravity:
  Gauss–Bonnet and trace extensions, Eur. Phys. J. C {\bfseries 76}, 578 
  (2016) \href{http://xxx.lanl.gov/abs/1606.05557}{{\ttfamily arXiv:1606.05557}}.

\bibitem[Bahamonde and Capozziello(2017)]{Bahamonde:2016grb}
S.~Bahamonde and S.~Capozziello, Noether Symmetry Approach in $f(T,B)$
  teleparallel cosmology, Eur. Phys. J. C {\bfseries 77}, 107 (2017) 
 \href{http://xxx.lanl.gov/abs/1612.01299}{{\ttfamily arXiv:1612.01299}}.

\bibitem[Golovnev et~al.(2017)Golovnev, Koivisto, and
  Sandstad]{Golovnev:2017dox}
A.~Golovnev, T.~Koivisto, and M.~Sandstad, On the covariance of teleparallel gravity theories, Classical Quantum Gravity {\bfseries 34}, 145013 (2017) 
 \href{http://xxx.lanl.gov/abs/1701.06271}{{\ttfamily arXiv:1701.06271}}.

\bibitem[Bahamonde et~al.(2017)Bahamonde, Böhmer, and
  Krššák]{Bahamonde:2017wwk}
S.~Bahamonde, C.~G. Böhmer, and M.~Krššák, New classes of modified
  teleparallel gravity models, Phys. Lett. B {\bfseries 775}, 37 (2017)
 \href{http://xxx.lanl.gov/abs/1706.04920}{{\ttfamily arXiv:1706.04920}}.

\bibitem[Mai and Lu(2017)]{Mai:2017riq}
Z.-F. Mai and H.~Lu, Black Holes, Dark Wormholes and Solitons in $f(T)$
  Gravities, Phys. Rev. D {\bfseries 95}, 124024 (2017) 
 \href{http://xxx.lanl.gov/abs/1704.05919}{{\ttfamily arXiv:1704.05919}}.

\bibitem[Hohmann et~al.(2018)Hohmann, Järv, and Ualikhanova]{Hohmann:2018rwf}
M.~Hohmann, L.~Järv, and U.~Ualikhanova, Covariant formulation of
  scalar-torsion gravity, Phys. Rev. D {\bfseries 97}, 104011 (2018)  \href{http://xxx.lanl.gov/abs/1801.05786}{{\ttfamily arXiv:1801.05786}}.

\bibitem[Ferraro and Fiorini(2008)]{Ferraro:2008ey}
R.~Ferraro and F.~Fiorini, On Born-Infeld Gravity in Weitzenbock
  spacetime, Phys. Rev. D {\bfseries 78}, 124019 (2008) \href{http://xxx.lanl.gov/abs/0812.1981}{{\ttfamily arXiv:0812.1981}}.

\bibitem[Bengochea and Ferraro(2009)]{Bengochea:2008gz}
G.~R. Bengochea and R.~Ferraro, Dark torsion as the cosmic speed-up, Phys. Rev. D {\bfseries 79}, 124019 (2009) \href{http://xxx.lanl.gov/abs/0812.1205}{{\ttfamily arXiv:0812.1205}}.

\bibitem[Wright(2016)]{Wright:2016ayu}
M.~Wright, Conformal transformations in modified teleparallel theories of
  gravity revisited, Phys. Rev. D {\bfseries 93}, 103002 (2016) \href{http://xxx.lanl.gov/abs/1602.05764}{{\ttfamily arXiv:1602.05764}}.

\bibitem[Ferraro and Fiorini(2015)]{Ferraro:2014owa}
R.~Ferraro and F.~Fiorini, Remnant group of local Lorentz transformations in
  $f(T)$ theories, Phys. Rev. D {\bfseries 91}, 064019 (2015) \href{http://xxx.lanl.gov/abs/1412.3424}{{\ttfamily arXiv:1412.3424}}.

\bibitem[Bejarano et~al.(2015)Bejarano, Ferraro, and Guzmán]{Bejarano:2014bca}
C.~Bejarano, R.~Ferraro, and M.~J. Guzmán, Kerr geometry in $f(T)$
  gravity, Eur. Phys. J. C {\bfseries 75}, 77 (2015) \href{http://xxx.lanl.gov/abs/1412.0641}{{\ttfamily arXiv:1412.0641}}.

\bibitem[Bejarano et~al.(2017)Bejarano, Ferraro, and Guzmán]{Bejarano:2017akj}
C.~Bejarano, R.~Ferraro, and M.~J. Guzmán, McVittie solution in $f(T)$
  gravity, Eur. Phys. J. C {\bfseries 77}, 825 (2017) \href{http://xxx.lanl.gov/abs/1707.06637}{{\ttfamily arXiv:1707.06637}}.

\bibitem[Hayashi and Shirafuji(1979)]{Hayashi:1979qx}
K.~Hayashi and T.~Shirafuji, New general relativity, Phys. Rev. D {\bfseries 19}, 3524 (1979); Addendum to ``New general relativity'', Phys. Rev. D {\bfseries 24}, 3312 (1981).

\bibitem[Ort{\'\i}n(2007)]{Ortin2007}
T.~Ort{\'\i}n, \textit{Gravity and strings} (Cambridge University Press, Cambridge, England, 2007).

\bibitem[Oshita and Wu(2017)]{Oshita:2017nhn}
N.~Oshita and Y.-P. Wu, Role of spacetime boundaries in a vierbein formulation of gravity, Phys. Rev. D {\bfseries 96}, 044042 (2017) \href{http://xxx.lanl.gov/abs/1705.10436}{{\ttfamily arXiv:1705.10436}}.

\bibitem[Anderson and Bergmann(1951)]{Anderson:1951ta}
J.~L. Anderson and P.~G. Bergmann, Constraints in covariant field
  theories, Phys. Rev. {\bfseries 83}, 1018 (1951).

\bibitem[Dirac(1964)]{Dirac1964}
P.~A.~M. Dirac, \textit{Lectures on quantum mechanics}, Belfer Graduate
  School of Science, (Yeshiva University, New York, 1964).

\bibitem[Sundermeyer(1982)]{Sundermeyer:1982gv}
K.~Sundermeyer, Constrained dynamics with applications to Yang-Mills theory,
  general relativity, classical spin, dual string model, Lect. Notes
  Phys. {\bfseries 169}, 1 (1982).

\bibitem[Sundermeyer(2014)]{Sundermeyer:2014kha}
K.~Sundermeyer, \textit{Symmetries in fundamental physics}, (Springer, Cham,
  Switzerland, 2014).

\bibitem[Henneaux and Teitelboim(1992)]{Henneaux:1992ig}
M.~Henneaux and C.~Teitelboim, \textit{Quantization of gauge systems} (University Press, Princeton, 1992).

\bibitem[Ferraro and Guzmán(2016)]{Ferraro:2016wht}
R.~Ferraro and M.~J. Guzmán, Hamiltonian formulation of teleparallel
  gravity, Phys. Rev. D {\bfseries 94}, 104045 (2016) \href{http://xxx.lanl.gov/abs/1609.06766}{{\ttfamily arXiv:1609.06766}}.

\bibitem[Krššák and Saridakis(2016)]{Krssak:2015oua}
M.~Krššák and E.~N. Saridakis, The covariant formulation of $f(T)$
  gravity, Classical Quantum Gravity {\bfseries 33}, 115009 (2016) \href{http://xxx.lanl.gov/abs/1510.08432}{{\ttfamily arXiv:1510.08432}}.

\bibitem[Nester and Ong(2017)]{Nester:2017wau}
J.~M. Nester and Y.~C. Ong, Counting components in the Lagrange multiplier
  formulation of teleparallel theories, \href{http://xxx.lanl.gov/abs/1709.00068}{{\ttfamily arXiv:1709.00068}}.

\bibitem[Maluf and da~Rocha-Neto(1999)]{Maluf:1998ae}
J.~W. Maluf and J.~F. da~Rocha-Neto, General relativity on a null surface:
  Hamiltonian formulation in the teleparallel geometry, Gen. Relativ. Gravit.  {\bfseries 31}, 173 (1999) \href{http://xxx.lanl.gov/abs/gr-qc/9808001}{{\ttfamily arXiv:gr-qc/9808001}}.

\bibitem[Maluf and da~Rocha-Neto(2001)]{Maluf:2001rg}
J.~W. Maluf and J.~F. da~Rocha-Neto, Hamiltonian formulation of general
  relativity in the teleparallel geometry, Phys. Rev. D {\bfseries 64}, 084014 
  (2001).

\bibitem[da~Rocha~Neto et~al.(2010)da~Rocha~Neto, Maluf, and
  Ulhoa]{daRochaNeto:2011ir}
J.~F. da~Rocha~Neto, J.~W. Maluf, and S.~C. Ulhoa, Hamiltonian formulation
  of unimodular gravity in the teleparallel geometry, Phys. Rev. D {\bfseries 82}, 124035 (2010)  \href{http://xxx.lanl.gov/abs/1101.2425}{{\ttfamily arXiv:1101.2425}},
[Erratum: Phys. Rev. D {\bfseries 87}, 069909 (2013)].

\bibitem[Ferraro and Guzmán(2018)]{Ferraro2018b}
R.~Ferraro and M.~J. Guzmán (to be published).

\end{thebibliography}
\end{document}